\documentclass{nseJournal}

\usepackage[square,sort,comma,numbers]{natbib}

\usepackage{lineno,hyperref}
\usepackage{natbib}
\usepackage{geometry}
\usepackage{fleqn}
\usepackage{graphicx}
\usepackage{txfonts}
\usepackage{hyperref}
\usepackage{booktabs}

\usepackage{bm}
\usepackage{amsmath}
\usepackage{amssymb}
\usepackage[justification=centering]{caption}
\modulolinenumbers[5]

\usepackage{framed} 
\usepackage{nomencl} 
\makenomenclature

\usepackage{color,soul}
\usepackage{subcaption}

\usepackage{multicol}
\usepackage{enumitem}

\usepackage{mathtools}    
\usepackage{multirow}    
\usepackage{subcaption}    
\usepackage{bm}    

\newcommand {\pp}[1] {\left( #1 \right)}

\newcommand {\brk}[1] {\left[ #1 \right]}

\newcommand {\pd} {\partial}    
\newcommand {\T} {^{\mathsf{T}}}    
\newcommand {\bu}[1] {\mathbf{#1}}    
\newcommand {\cg}[1] {\mathcal{#1}}    
\newcommand {\bb}[1] {\mathbb{1}}    

\renewcommand {\hat}[1] {\widehat{#1}}    

\newcommand {\btheta} {\bm{\theta}}
\newcommand {\bmu} {\bm{\mu}}

\newcommand {\yE} {\bu{y}_\mathrm{E}}
\newcommand {\yR} {\bu{y}_\mathrm{R}}
\newcommand {\yM} {\bu{y}_\mathrm{M}}
\newcommand {\yA} {\bu{y}_\mathrm{A}}
\newcommand {\yB} {\bu{y}_\mathrm{B}}

\newcommand {\yMpost} {\hat{\bu{y}}_\mathrm{M}}
\newcommand {\yApost} {\hat{\bu{y}}_\mathrm{A}}
\newcommand {\yBpost} {\hat{\bu{y}}_\mathrm{B}}

\newcommand {\SigmaE} {\bm{\Sigma}_\mathrm{E}}
\newcommand {\SigmaM} {\bm{\Sigma}_\mathrm{M}}
\newcommand {\SigmaA} {\bm{\Sigma}_\mathrm{A}}
\newcommand {\SigmaB} {\bm{\Sigma}_\mathrm{B}}
\newcommand {\SigmaAB} {\bm{\Sigma}_\mathrm{AB}}
\newcommand {\SigmaTh} {\bm{\Sigma}_{\btheta}}
\newcommand {\SigmaThB} {\bm{\Sigma}_{\btheta \mathrm{B}}}

\newcommand {\SigmaMpost} {\hat{\bm{\Sigma}}_\mathrm{M}}
\newcommand {\SigmaApost} {\hat{\bm{\Sigma}}_\mathrm{A}}
\newcommand {\SigmaBpost} {\hat{\bm{\Sigma}}_\mathrm{B}}
\newcommand {\SigmaABpost} {\hat{\bm{\Sigma}}_\mathrm{AB}}
\newcommand {\SigmaThpost} {\hat{\bm{\Sigma}}_{\btheta}}

\newcommand {\sA} {\bm{S}_\mathrm{A}}
\newcommand {\sB} {\bm{S}_\mathrm{B}}

\newcommand {\kMeff} {k^\mathrm{M}_\text{eff}}

\begin{document}

\title{Nuclear Data Adjustment for Nonlinear Applications in the OECD/NEA WPNCS SG14 Benchmark - A Bayesian Inverse UQ-based Approach for Data Assimilation}

\addAuthor{Christopher Brady}{a}
\addAuthor{\correspondingAuthor{Xu Wu}}{a}
\correspondingEmail{clbrady@ncsu.edu}

\addAffiliation{a}{Department of Nuclear Engineering, North Carolina State University    \\ 
	Burlington Engineering Laboratories, 2500 Stinson Drive, Raleigh, NC 27695 \\}


\addKeyword{Inverse Uncertainty Quantification}
\addKeyword{Bayesian Calibration}
\addKeyword{Nuclear Data Adjustment}
\addKeyword{Data Assimilation}
\addKeyword{Scientific Machine Learning}

\titlePage

\begin{abstract}
The Organization for Economic Cooperation and Development (OECD) Working Party on Nuclear Criticality Safety (WPNCS) proposed a benchmark exercise to assess the performance of current nuclear data adjustment techniques applied to nonlinear applications and experiments with low correlation to applications. This work introduces Bayesian Inverse Uncertainty Quantification (IUQ) as a method for nuclear data adjustments in this benchmark, and compares IUQ to the more traditional methods of Generalized Linear Least Squares (GLLS) and Monte Carlo Bayes (MOCABA). Posterior predictions from IUQ showed agreement with GLLS and MOCABA for linear applications. When comparing GLLS, MOCABA, and IUQ posterior predictions to computed model responses using adjusted parameters, we observe that GLLS predictions fail to replicate computed response distributions for nonlinear applications, while MOCABA shows near agreement, and IUQ uses computed model responses directly. We also discuss observations on why experiments with low correlation to applications can be informative to nuclear data adjustments and identify some properties useful in selecting experiments for inclusion in nuclear data adjustment. Performance in this benchmark indicates potential for Bayesian IUQ in nuclear data adjustments.
\end{abstract}

\section{Introduction}

Advanced and small modular reactors employ unseasoned designs often using higher enriched fuel which introduces regulatory, safeguards, and optimization challenges. Meeting these challenges ensure nuclear technologies are safe to the public, secure from proliferation in a dynamic political climate, and economically feasible in a transitioning power grid. Increased model accuracies will assist in meeting regulatory and optimization requirements while highly sensitive detection techniques enable proliferation countermeasures. Nuclear data adjustments provide a method of reducing model prediction uncertainties caused by uncertainties in the nuclear data. Improving the accuracy in nuclear data improves our ability to leverage advanced computing and simulations in the pursuit of these endeavors. Several efforts are underway to assess current data adjustment methodologies, make recommendations on best practices, or improve on current methodologies~\cite{saintjean_2011_assessment, salvatores_2014_methods, siefman_2018_stochastic}.

In the broader context, \textit{data assimilation} (DA) is described as a process to combine observational data with a model to improve the model predictions and is broadly rooted in the areas of weather forecasting and geosciences~\cite{asch_2016_data, reich_2015_probabilistic, carrassi_2018_data}. Generally, DA seeks to determine the system's state, initial or boundary conditions, or other model parameters from measured or observed data. Inferring the true but unknown values of model parameters from observed data is inherently an \textit{inverse problem} and the general way to describe the parameter space is to define the parameter probability densities~\cite{tarantola_2005_inverse}. Inverse problems naturally arise in many fields of research where knowledge of the causal factors is useful but not directly available, while responses from these causal factors are readily observable. Similar inverse problems such as calibration, data adjustment, data-informed modeling, error recovery, inverse uncertainty quantification (UQ), and potentially others either overlap or are synonymous with DA. Bayesian inference is well suited to framing the inverse problem, combining the available \textit{a priori} information, and producing \textit{a posteriori} parameter densities. These posterior parameter densities provide analysts with a robust description of the parameter densities which enhances the propagation of uncertainty through models that rely on these parameters to make predictions.

Possibly the most prevalent method for nuclear data adjustment is generalized linear least squares (GLLS)~\cite{pazy_1974_role, saintjean_2011_assessment, salvatores_2014_methods, broadhead_2004_sensitivity, rearden_2011_sensitivity, kiedrowski_2015_whisper}. This method has proven to be effective in reducing application uncertainties in order to establish safe upper subcritical limits in criticality safety applications for decades. While this method is pragmatic and demonstrated in real applications, several drawbacks are identified. Primarily, GLLS relies on the assumption that the application response is linear over the domain of the uncertain parameters. While this assumption is often valid for the purposes of criticality safety, it does not hold for many other applications that do not have linear responses. Furthermore, this method assumes applications are normally distributed and only provides information on the expectation and covariance of the posterior distributions without regard for skewness or kurtosis.

Some of these issues have been addressed with methods such as Monte Carlo-Bayes (MOCABA)~\cite{watanabe_2014_cross, hoefer_2015_mocaba, hoefer_2019_applications, hoefer_2021_assessing} or Bayesian Monte Carlo (BMC)~\cite{capote_2008_investigation, saintjean_2018_evaluation}. Basic MOCABA avoids the linearity assumptions by evaluating the models at samples drawn from the prior parameter distribution to produce sample covariance estimates instead of using linear transformations. Generalized MOCABA avoids the normality assumption by transforming the sampled model evaluations to near normal distributions for the data adjustment then inversely transforming the results to the posterior predictive distributions. Where GLLS can be viewed as analogous to a Kalman Filter, basic MOCABA can be viewed as analogous to an Ensemble Kalman Filter~\cite{mandel_2009_brief, kalman_1960_new}. BMC also avoids the linearity assumptions by evaluating the models at samples drawn from the prior parameter distribution. However, BMC uses these samples to evaluate the likelihood of data given the model response at each parameter sample. These likelihood evaluations form the weights used to compute the weighted mean estimators of the mean vector and covariance matrix of the posterior parameter and posterior predictive distributions. Because posterior distributions are defined only by the mean and covariance, no information on skewness or kurtosis is provided by this method. BMC has been shown to have general agreement with GLLS and basic MOCABA~\cite{siefman_2018_stochastic}. Because BMC produces similar results to MOCABA but does not inform higher moments in the posterior, this work will not include BMC in the comparative study.

The main objective of this work is to adopt and improve a Bayesian Inverse Uncertainty Quantification (IUQ) approach that has been widely used for statistical model calibration to a nuclear data adjustment problem. Bayesian IUQ was originally developed and demonstrated in system nuclear thermal-hydraulics~\cite{wu_2018_inverse1, wu_2018_inverse2} and nuclear fuel performance problems~\cite{wu_2018_krigingbased}. We will apply this method to a well-defined benchmark exercise, and compare our results with those produced by GLLS and MOCABA. While IUQ has its challenges, it provides complete empirical posterior parameter distributions, is not restricted by the same linearity and normality assumptions, and can be adjusted to match our confidence in prior information. However, these benefits come with a higher computational cost, which is often addressed through surrogate modeling. It should be reinforced that this work is not meant to suggest a problem with the GLLS methodology as it has been effectively applied in nuclear criticality safety, but to explore how effectively other methodologies might expand nuclear data adjustment to applications with nonlinear behaviors.

The Organization for Economic Cooperation and Development (OECD) Nuclear Energy Agency (NEA) Working Party for Nuclear Criticality Safety (WPNCS) Sub-Group 14 (SG14) created a performance benchmark for error recovery and experimental coverage. The stated objective of the benchmark is to build confidence in DA techniques and adjusted nuclear data under existing and anticipated challenging conditions. This performance benchmark was designed with the intent to have a wide variety of applications including:  criticality safety, spent fuel characterization, online reactor monitoring, and safeguards. We use this benchmark exercise as a test case for the Bayesian IUQ process applied to nuclear data adjustment. The details of the benchmark exercise are described in Section~\ref{section:Problem-Pefinition}.

The results of this exercise indicate general agreement between GLLS, MOCABA, and IUQ for linear applications. While GLLS posterior predictions fail for nonliner applications, MOCABA and IUQ show promise in capturing higher order moments in posterior predictions. When comparing the nonlinear applications' posterior predictive distributions to the distributions of computed model responses using posterior parameter samples, we observe disagreement for GLLS, near agreement in MOCABA, and agreement in IUQ as the computed model responses directly form the posterior predictive distribution. Data adjustments using nonlinear experiments remain untested in this exercise. Some properties indicating experimental relevance to a data adjustment are identified and discussed along with the effectiveness of the correlation coefficient as a metric of experimental relevance. Based on the results of this exercise, IUQ has strong potential to perform nuclear data adjustments in nonlinear applications. The method's ability to  scale to higher dimension parameter spaces and overcome computational requirements must be demonstrated, but approaches to address these concerns are available.

The remainder of this article is arranged as follows: Section~\ref{section:Problem-Pefinition} defines the problem proposed by the performance benchmark; Section~\ref{section:Methodologies} outlines the GLLS, MOCABA, and Bayesian IUQ methodologies; Section~\ref{section:Results} presents the results and discusses the findings; and Section~\ref{section:Conclusions} summarizes this work and highlights key observations.

\section{Problem Definition}
\label{section:Problem-Pefinition}

The OECD/NEA WPNCS SG14 benchmark problem consists of 4 \textsc{Experiments} (\texttt{Albert}, \texttt{Bohr}, \texttt{Chadwick}, and \texttt{Dyson}) and 3 \textsc{Applications} (\texttt{Bravo}, \texttt{Castle}, and \texttt{Trinity}) where the response emulates the behavior of critical eigenvalues at a range of operating conditions. The various operating conditions, such as fuel-to-moderator ratios, in each of the \textsc{Experiments} and \textsc{Applications} yields a range of replicated neutronic behaviors. Specifically, some \textsc{Applications} have nonlinear responses over the parameter domain and some \textsc{Experiments} exhibit small or negative correlation with the \textsc{Applications}. Additionally, error sources unknown to participants are embedded into the models to assess and compare the performance of each proposed DA technique. Participants are tasked with providing adjusted parameters (i.e., posterior parameter distributions), predicted \textsc{Application} responses with uncertainty (i.e., posterior predictive distributions), response sensitivity to parameters, the ability to quantify experimental relevance, and any additional results from the chosen methodology. Methods are to be assessed by their ability to identify and adjust for embedded error sources and to accurately predict \textsc{Application} bias and uncertainty.

For each of the 4 \textsc{Experiments} and 3 \textsc{Applications}, participants are provided a computational model $\yM$ to compute the effective multiplication factor $\kMeff$ (i.e. critical eigenvalue). The model takes 5 input parameters $\btheta$ representing nuclear cross-sections.
\begin{equation}
    \yM \pp{\btheta} = \kMeff
\end{equation}
The prior parameter distribution is meant to represent nuclear covariance data that could be provided by a standard nuclear data library and is distributed as multivariate normal $\btheta \sim \cg{N}_5 \pp{\bm{\mu}_{\btheta}, \SigmaTh}$ shown in expanded form in Table~\ref{tab:prior} with values truncated to three digits for space.

\begin{table}[!htb]
    \centering
    \caption{Prior parameter distribution $\btheta \sim \cg{N}_5 \pp{ \bm{\mu}_{\btheta}, \SigmaTh }$}
    \vspace{-1em}
    \begin{equation*}
        \begin{bmatrix}
            \theta_1 \\ \theta_2 \\ \theta_3 \\ \theta_4 \\ \theta_5
        \end{bmatrix}
        \sim \cg{N}_5
        \begin{bmatrix}
            \begin{pmatrix*}[r]
                13.1 \\ 14.1 \\ 6.2 \\ 14.1 \\ 11.3
            \end{pmatrix*},
            \begin{pmatrix*}[r]
                1.00 & 0.17 & 0.07 & -0.02 & -0.02 \\
                0.17 & 0.77 & 0.19 & 0.06 & 0.22 \\
                0.07 & 0.19 & 0.94 & -0.03 & -0.07 \\
                -0.02 & 0.06 & -0.03 & 0.75 & 0.18 \\
                -0.02 & 0.22 & -0.07 & 0.18 & 0.23 \\
            \end{pmatrix*}
        \end{bmatrix}
    \end{equation*}
    \label{tab:prior}
\end{table}

Additionally, each \textsc{Experiment} and \textsc{Application} is accompanied by a nominal model response computed at the prior parameter mean value as $\yM \pp{\bm{\mu}_\theta}$ and an approximated response standard deviation. To approximate this response standard deviation, the model is assumed linear within the parameter domain around the nominal model response. This assumption implies that the model covariance $\SigmaM$ is simply a linear transformation of the parameter covariance $\SigmaTh$ using the collection of model gradients $\bm{S}$ evaluated at $\bm{\mu}_\theta$ as shown in Eqn.~(\ref{eqn:sandwich_rule}). The approximated model standard deviations are then extracted from the model covariance diagonal in the standard fashion.
\begin{equation}
    \bm{\Sigma}^\mathrm{M} \approx \bm{S} \SigmaTh \bm{S}\T
    \qquad \text{where} \qquad 
    \bm{S} =
    \begin{bmatrix}
        S_{11} & \dots  & S_{1p} \\
        \vdots & \ddots & \vdots \\
        S_{n1} & \dots  & S_{np}
    \end{bmatrix}
    \qquad
    \text{and}
    \qquad
    S_{ij} = \frac{\pd k^\mathrm{M,i}_\text{eff}}{\pd \theta_j} \bigg\rvert_{\theta_{j} = \mu_{\theta, j}} 
    \label{eqn:sandwich_rule}
\end{equation}
Here $i=1,\dots,n$ where $n$ is the number of models and $j=1,\dots,p$ where $p$ is the number of parameters. As the matrix $\bm{S}$ is not provided by the benchmark exercise, we approximate it using the forward difference method and reproduce the provided model standard deviations. For each \textsc{Experiment}, an experimentally measured or observed value $\yE$ is provided with accompanying measurement standard deviation. The provided nominal model responses and experimentally measured values with standard deviations are summarized with truncated values in Table~\ref{tab:responses}. 

\begin{table}[!htb]
    \centering
    \caption{Modeled and measured values of response $k_\text{eff}$ for each \textsc{Experiment} and \textsc{Application}.}
    \begin{tabular}{l l|c c|c c|}
        & & \multicolumn{2}{c|}{Modeled} & \multicolumn{2}{c|}{Measured} \\
        & & nominal & std & mean & std \\ \hline
        \multirow{4}{*}{\textsc{Experiments}}  & \texttt{Albert}   & 0.99061 & 0.01001 & 0.99938 & 0.001 \\
        & \texttt{Bohr}     & 0.99541 & 0.01002 & 1.00026 & 0.001 \\
        & \texttt{Chadwick} & 0.99359 & 0.00692 & 0.99525 & 0.001 \\
        & \texttt{Dyson}    & 0.98745 & 0.00701 & 0.99560 & 0.001 \\ \hline
        \multirow{3}{*}{\textsc{Applications}} & \texttt{Bravo}    & 0.99596 & 0.02898 & - & - \\
        & \texttt{Castle}   & 0.99422 & 0.01463 & - & - \\
        & \texttt{Trinity}  & 0.99711 & 0.03021 & - & - \\ \hline
    \end{tabular}
    \label{tab:responses}
\end{table}

Each \textsc{Experiment} and \textsc{Application} model also provides a normalized sensitivity profile $\bm{S}^N$ computed as the gradient of the model response with respect to the 5 input parameters evaluated at $\bm{\mu}_\theta$, then normalized by $\bm{\mu}_\theta$ and $\kMeff$ as shown in Eqn.~(\ref{eqn:normalized_sensitivity}). 
\begin{equation}
    \bm{S}_i^N =
    \begin{bmatrix}
        S_{i1}^N & \dots & S_{ip}^N
    \end{bmatrix}
    \quad \quad
    \text{where}
    \quad \quad
    S_{ij}^N = \frac{\pd k^\mathrm{M,i}_\text{eff}}{\pd \theta_j} \frac{\theta_j}{k^\mathrm{M,i}_\text{eff}} \bigg\rvert_{\theta_j = \mu_{\theta,j}} 
    \label{eqn:normalized_sensitivity}
\end{equation}
A normalized model gradient is provided because \textit{relative} parameter and response covariances are sometimes used in criticality applications~\cite{broadhead_2004_sensitivity, kiedrowski_2015_whisper}, though standard covariances are used throughout this exercise. The provided normalized sensitivity profiles are summarized with truncated values in Table~\ref{tab:sensitivity}. While these values are provided by the benchmark exercise, they are not used in this paper since the provided covariances are not \textit{relative}.

\begin{table}[!htb]
    \centering
    \caption{Normalized sensitivity profile of the response with respect to input parameters.}
    \begin{tabular}{l l|r r r r r|}
        & & \multicolumn{1}{c}{$S_1^N$} & \multicolumn{1}{c}{$S_2^N$} & \multicolumn{1}{c}{$S_3^N$} & \multicolumn{1}{c}{$S_4^N$} & \multicolumn{1}{c|}{$S_5^N$} \\ \hline
        \multirow{4}{*}{\textsc{Experiments}}  & \texttt{Albert}   & -0.019 & -0.096 & -0.008 & -0.086 & -0.041 \\
        & \texttt{Bohr}     & -0.011 & -0.103 & -0.009 & -0.086 & -0.031 \\
        & \texttt{Chadwick} &  0.053 &  0.010 &  0.020 & -0.052 & -0.028 \\
        & \texttt{Dyson}    &  0.003 & -0.059 &  0.008 & -0.062 & -0.047 \\ \hline
        \multirow{3}{*}{\textsc{Applications}} & \texttt{Bravo}    & -0.054 & -0.267 & -0.026 & -0.257 & -0.123 \\
        & \texttt{Castle}   & -0.028 & -0.139 & -0.011 & -0.128 & -0.059 \\
        & \texttt{Trinity}  & -0.051 & -0.309 & -0.025 & -0.239 & -0.119 \\ \hline
    \end{tabular}
    \label{tab:sensitivity}
\end{table}

Each \textsc{Experiment} is also provided its correlation coefficient $c_k$ with respect to each of the \textsc{Applications} outlined in Table~\ref{tab:correlation}. This is the Pearson correlation computed from the model covariance $\SigmaM$ as provided in Eqn.~(\ref{eqn:sandwich_rule}). In criticality applications, $c_k$ is often used as a measure of similarity between the \textsc{Experiment} and \textsc{Application} and sometimes used to justify inclusion or exclusion from a set of \textsc{Experiments} used to perform a nuclear data adjustment~\cite{broadhead_2004_sensitivity, kiedrowski_2015_whisper}.

\begin{table}[!htb]
    \centering
    \caption{Correlation of \textsc{Experiments} with \textsc{Applications} $c_k$.}
    \begin{tabular}{l l|r r r|}
        & & \multicolumn{3}{c|}{\textsc{Applications}} \\
        & & \texttt{Bravo}  & \texttt{Castle} & \texttt{Trinity}  \\ \hline
        \multirow{4}{*}{\textsc{Experiments}}  & \texttt{Albert} & 0.9996 & 0.9592 & 0.9986  \\
        & \texttt{Bohr}     & 0.9972 &  0.9932 & 0.9982  \\
        & \texttt{Chadwick} & 0.1073 & -0.1968 & 0.0721  \\
        & \texttt{Dyson}    & 0.9081 &  0.7567 & 0.8995  \\ \hline
    \end{tabular}
    \label{tab:correlation}
\end{table}

The benchmark exercise directs participants to consider 5 cases, where each case uses a subset of the available experimentally measured data to make data adjustments and posterior response predictions. Comparing the cases provides an opportunity to assess, and potentially quantify, the contribution of each \textsc{Experiment} to the posterior parameter distributions and posterior predictive distributions. This serves as the basis of evaluating experimental relevance to the data adjustment. The subset of \textsc{Experiments} to be considered in each case is summarized in Table~\ref{tab:cases}.

\begin{table}[!htb]
    \centering
    \caption{Subset of \textsc{Experiments} used to perform data adjustments in each case.}
    \begin{tabular}{l|c c c c|}
        & \texttt{Albert} & \texttt{Bohr} & \texttt{Chadwick} & \texttt{Dyson} \\ \hline
        Case 1 & \checkmark &   &   &    \\ \hline
        Case 2 & \checkmark & \checkmark &   &    \\ \hline
        Case 3 & \checkmark &   & \checkmark &    \\ \hline
        Case 4 & \checkmark &   &   & \checkmark  \\ \hline
        Case 5 & \checkmark & \checkmark & \checkmark & \checkmark  \\ \hline
    \end{tabular}
    \label{tab:cases}
\end{table}

\section{Methodologies}
\label{section:Methodologies}

In this section, we will first define a general Bayesian framework for inverse problems, and then briefly summarize the GLLS and MOCABA methodologies before presenting the proposed Bayesian IUQ methodology for nuclear data adjustments. A summary of the notation used in this paper is provided in Table~\ref{tab:notation}. 

\begin{table}[!htb]
    \centering
    \caption{Summary of notation.}
    \begin{tabular}{c|l|}
        Symbol & Description \\ \hline
        $\bu{y}$ & response or QoI \\
        $\btheta$ & uncertain input parameters \\
        $\bu{x}$ & design variables \\
        $\bm{\varepsilon}$ & measurement errors \\
        $\bm{\Sigma}$ & covariance matrix \\
        $\bu{S}$ & model gradient at nominal parameters \\
        $\mathrm{R}$ & subscript identifying reality  \\
        $\mathrm{E}$ & subscript identifying experimentally measured data \\
        $\mathrm{M}$ & subscript identifying all models \\
        $\mathrm{A}$ & subscript identifying \textsc{Application} models \\
        $\mathrm{B}$ & subscript identifying \textsc{Experiment} models \\
        $\hat{\text{\;\;\;}}$ & modifier identifying posterior information \\
        \hline
    \end{tabular}
    \label{tab:notation}
\end{table}

\subsection{Bayesian Framework for the Inverse Problem}
\label{subsection:Bayes}

All three methodologies examined in this paper use Bayesian inference to approach the inverse problem. We typically frame the inverse problem involving a model with experimental data as follows:
\begin{equation}
    \yE(\bu{x}) = \yM(\bu{x},\btheta) + \bm{\varepsilon}
    \label{eqn:inv_model}
\end{equation}
where $\yE$ is the experimental observation, $\yM$ is the model response, $\bu{x}$ is the vector of design variables, $\btheta$ is the vector of model input parameters to be adjusted or calibrated, and $\bm{\varepsilon}$ is the vector of measurement errors. The design variables represent elements of the operating conditions of each of the \textsc{Experiments} or \textsc{Applications}. In this benchmark exercise, the design variables are obscured to us and entirely contained within each of the models, leaving only the input parameters for consideration.

The goal of Bayesian inference is to determine the posterior distribution of the input parameters given the model and data using Bayes Theorem in Eqn.~(\ref{eqn:bayes_theorem})
\begin{equation}
    \text{p}\pp{\btheta | \yE,\yM} = 
    \frac{ \text{p}\pp{\yE,\yM | \btheta} \cdot \text{p}\pp{\btheta} }{ \text{p}\pp{\yE,\yM} } =
    \frac{ \text{p}\pp{\yE,\yM | \btheta} \cdot \text{p}\pp{\btheta} }
    { \int \text{p}\pp{\yE,\yM | \btheta} \cdot \text{p}\pp{\btheta} \cdot d\btheta }
    \label{eqn:bayes_theorem}
\end{equation}
where $\text{p}\pp{\btheta | \yE,\yM}$ is the posterior, $\text{p}\pp{\yE,\yM | \btheta}$ is the likelihood, $\text{p}\pp{\btheta}$ is the prior, and $\int \text{p}\pp{\yE,\yM | \btheta} \cdot \text{p}\pp{\btheta} \cdot d\btheta$ is the marginal likelihood, also referred to as the evidence term or simply the normalizing constant.

\subsection{Generalized Linear Least Squares (GLLS)}
\label{subsection:GLLS}

The nuclear industry has safely performed nuclear data adjustments in nuclear criticality safety applications with GLLS for decades. This methodology is incorporated into prominent nuclear codes including SCALE~\cite{rearden_2011_sensitivity} and MCNP's Whisper~\cite{kiedrowski_2015_whisper}, among others. This methodology is effective for applications where the linearity and normality assumptions hold. For example, when the parameter uncertainties are small enough that the model response is effectively linear over the small domain of the uncertain parameters, which is true in many criticality safety scenarios. However, when the model exhibits strong nonlinearities over the domain of the uncertain parameters, we expect GLLS to yield erroneous parameter adjustments and response predictions. The goal of this benchmark is not to suggest that GLLS is somehow insufficient, but to explore if other methodologies might expand DA to applications beyond the current linearity limitations.

Many detailed descriptions of GLLS and its derivation are available~\cite{kalman_1960_new, pazy_1974_role, salvatores_2014_methods, saintjean_2011_assessment} with one very succinct and clear summary given in~\cite{watanabe_2014_cross}. In this paper, we will provide a brief mathematical description of GLLS to make this paper self-contained and to facilitate the comparison with MOCABA and IUQ. Starting with Bayes Theorem in Eqn.~(\ref{eqn:bayes_theorem}), GLLS assumes the prior parameters are distributed multivariate normal about the parameter nominal values, the likelihood of the data given the parameters is distributed multivariate normal about the model, and the marginal is distributed multivariate normal about the true value of the experimental data. A first-order approximation of the Taylor expansion is used as a linear approximation of the model to simplify computation of the likelihood derivative during likelihood maximization. This results in a maximum \textit{a posteriori} estimate of the posterior parameter distribution, or adjusted parameters, in terms of adjusted parameter mean and covariance shown in Eqn.~(\ref{eqn:glls_param}).
\begin{equation}
    \begin{aligned}
        \hat{\btheta} &= \btheta + \SigmaTh \sB\T \pp{ \sB \SigmaTh \sB\T + \SigmaE }^{-1} \pp{ \yE - \yB } \\
        \SigmaThpost &= \SigmaTh - \SigmaTh \sB\T \pp{ \sB \SigmaTh \sB\T + \SigmaE }^{-1} \sB \SigmaTh
    \end{aligned}
    \label{eqn:glls_param}
\end{equation}
The GLLS posterior predictions, or posterior responses, are then computed as linear transformations of the prior predictions using these adjusted parameters as shown in Eqn.~(\ref{eqn:glls_response}).
\begin{equation}
    \begin{aligned}
        \yApost = \yA + \sA \pp{ \hat{\btheta} - \btheta}&, \quad \yBpost = \yB + \sB \pp{ \hat{\btheta} - \btheta} \\
        \SigmaApost = \sA \SigmaThpost \sA\T, \quad
        \SigmaBpost = &\sB \SigmaThpost \sB\T, \quad
        \SigmaABpost = \sA \SigmaThpost \sB\T \\
    \end{aligned}
    \label{eqn:glls_response}
\end{equation}
The linear approximations will lead to errors when either the \textsc{Experiments} or the \textsc{Applications} have nonlinear behaviors over the parameter domain. As discussed in Section~\ref{subsection:Prior-Model-Behavior}, the \textsc{Experiments} involved in this benchmark exercise are linear, but two of the three \textsc{Applications} are nonlinear. Examining these equations implies GLLS should provide reasonable parameter adjustments because these are only a function of the linear \textsc{Experiments}. It should also provide reasonable posterior predictions for \texttt{Castle} which also behaves linearly. However, this formulation implies the posterior predictions will be normally distributed when nonlinear \textsc{Applications} might exhibit skewed distributions. Thus, GLLS is expected to yield poor posterior predictions for the nonlinear \textsc{Applications} \texttt{Bravo} and \texttt{Trinity}.

\subsection{Monte Carlo-Bayes (MOCABA)}
\label{subsection:MOCABA}

The MOCABA method was introduced to avoid the underlying linearity and normality assumptions that restrict GLLS. A thorough description of its implementation and derivation are available~\cite{hoefer_2015_mocaba, hoefer_2019_applications, hoefer_2021_assessing}, while a relatively succinct and intuitive derivation from the GLLS equations can be found in~\cite{watanabe_2014_cross}. Similar to GLLS, in this paper we will only briefly describe the MOCABA framework. MOCABA organizes the \textsc{Applications} and \textsc{Experiments} according to Eqn.~(\ref{eqn:mocaba_structure}).
\begin{equation}
    \yM = 
    \begin{pmatrix}
        \yA \\ \yB
    \end{pmatrix}
    , \qquad
    \SigmaM = 
    \begin{pmatrix}
        \SigmaA & \SigmaAB  \\
        \SigmaAB\T & \SigmaB  \\
    \end{pmatrix}
    \label{eqn:mocaba_structure}
\end{equation}
Values for these are estimated by drawing parameter samples from the prior parameter distribution and computing model responses for each parameter sample. Using these computed response samples, $\yM$ is estimated as the sample mean, $\SigmaM$ is estimated as the sample covariance, and $\SigmaThB$ is estimated as the sample cross covariance between the parameters and the \textsc{Experiments}. MOCABA circumvents the necessity of using local sensitivities to perform linear transformations by replacing $\sB \SigmaTh \sB\T$ and $\SigmaTh \sB\T$ in Eqn.~(\ref{eqn:glls_param}) with $\SigmaB$ and $\SigmaThB$ respectively. This provides intuition for the posterior parameter distribution estimates in Eqn.~(\ref{eqn:mocaba_param}).
\begin{equation}
    \begin{aligned}
        \hat{\btheta} &= \btheta + \SigmaThB \pp{ \SigmaB + \SigmaE }^{-1} \pp{ \yE - \yB } \\
        \SigmaThpost &= \SigmaTh - \SigmaThB \pp{ \SigmaB + \SigmaE }^{-1} \SigmaThB
    \end{aligned}
    \label{eqn:mocaba_param}
\end{equation}
The posterior prediction estimates in Eqn.~(\ref{eqn:mocaba_response}) are derived through maximum likelihood estimation on the following slightly modified framing of Bayes theorem $p\pp{\yM | \yE} \propto  p\pp{\yE | \yM} \cdot p\pp{\yM}$.
\begin{equation}
    \begin{aligned}
        \yApost &= \yA + \SigmaB \pp{ \SigmaB + \SigmaE }^{-1} \pp{ \yE - \yB } \\
        \yBpost &= \yB + \SigmaB \pp{ \SigmaB + \SigmaE }^{-1} \pp{ \yE - \yB } \\
        \SigmaApost &= \SigmaA - \SigmaAB \pp{ \SigmaB + \SigmaE }^{-1} \SigmaAB\T \\
        \SigmaBpost &= \SigmaB - \SigmaB \pp{ \SigmaB + \SigmaE }^{-1} \SigmaB \\
        \SigmaABpost &= \SigmaAB - \SigmaAB \pp{ \SigmaB + \SigmaE }^{-1} \SigmaB
    \end{aligned}
    \label{eqn:mocaba_response}
\end{equation}

While MOCABA still assumes normally distributed responses during the parameter adjustments and posterior prediction estimates, the limitation is reduced by transforming the sampled prior response distribution to a normal distribution before analysis, then inversely transforming the posterior response distribution following the analysis. To achieve this, we use an empirical quantile transform on the Monte Carlo response samples prior to the data adjustment, then perform the inverse quantile transform on samples drawn from the normal posterior predictive distribution to obtain samples from the posterior predictive distribution.

Another method of obtaining samples of posterior predictions following inverse transformation is suggested~\cite{hoefer_2021_assessing} where data variance is included in the prior response sampling and thus contained in $\SigmaB$ making $\SigmaE$ become 0. This is achieved by drawing samples of $\bm{\varepsilon}$ from $\cg{N} \pp{0, \SigmaE}$ then adding them to the response samples. Without this step, it is necessary to transform $\SigmaE$ in the same fashion as the response samples are transformed to a normal distribution. Performing such a transformation is difficult as it is inherently a nonlinear transformation. In this method posterior samples are obtained:
\begin{equation}
    \hat{\bu{y}}_{\mathrm{M},j} = \yMpost + \hat{\bu{L}} \bu{L}^{-1}  \pp{\bu{y}_{\mathrm{M},j} + \bm{\varepsilon}_j - \yM}, \quad j=1, \dots, m
    \label{eqn:mocaba_posterior_samples}
\end{equation}
where $m$ is the number of samples and $\hat{\bu{L}}$ and $\bu{L}$ are obtained from the Cholesky decomposition:
\begin{equation}
    \SigmaM = \bu{L} \bu{L} \T, \qquad \SigmaMpost = \hat{\bu{L}} \hat{\bu{L}}\T
    \label{eqn:mocaba_cholesky1}
\end{equation}
However, when $\SigmaE$ is 0 an identity matrix arises in computing $\SigmaBpost$ and $\SigmaABpost$ in Eqn.~(\ref{eqn:mocaba_response}) causing these estimates to be 0, in turn making Cholesky decomposition of $\SigmaMpost$ impossible. An alternative suggestion~\cite{hoefer_2021_assessing} is to estimate the transformed $\bm{\Sigma}_{z \mathrm{E}}$ with a linear approximation using the derivative of the transformation. Instead we use Cholesky decomposition on the covariance of the sampled responses and the transformed responses to build a scaling matrix used to estimate the transformed $\SigmaE$.
\begin{equation}
    \bm{\Sigma}_{z \mathrm{E}} = \pp{\bu{L}_z \bu{L}_\mathrm{B}^{-1}} \SigmaE \pp{\bu{L}_z \bu{L}_\mathrm{B}^{-1}}\T
    \label{eqn:mocaba_transformation}
\end{equation}
where $z$ indicates the transformed response distribution and Cholesky decomposition is performed:
\begin{equation}
    \SigmaB = \bu{L}_\mathrm{B} \bu{L}_\mathrm{B} \T, \qquad \bm{\Sigma}_z = \bu{L}_z \bu{L}_z^{\mathsf{T}}
    \label{eqn:mocaba_cholesky2}
\end{equation}
Both the linear approximation and the scaling matrix are linear estimates of a nonlinear transformation and will thus result in some warping of the experimentally measured covariance.

\subsection{Bayesian IUQ}
\label{subsection:IUQ}

A distinction should be made between the often more familiar \textit{forward} UQ (FUQ) process and the \textit{inverse} UQ (IUQ) process~\cite{wu_2021_comprehensive}. FUQ propagates parameter uncertainties through a model to inform uncertainties in a QoI. However, FUQ requires well defined input parameter uncertainties, which are often not available or are subjective. In the case of nuclear cross section data, the evaluated nuclear data files from various nuclear data libraries do not always agree and the evaluation process involves some subjectivity, which can reduce confidence in our model predictions and are problematic for regulatory or law enforcement purposes. Conversely, IUQ takes experimental data on a QoI with measurement uncertainty and propagates the uncertainties through the model to inform the uncertainties in the input parameters. Using statistical methods, the IUQ process provides mathematical rigor to quantify uncertainties in the input parameters given the model and experimental data.

In modeling and simulation, we typically consider five quantifiable sources of uncertainty: parameter, experimental, numerical, model, and code~\cite{wu_2018_inverse1, wu_2018_inverse2}. \textit{Parameter uncertainty} in the model input parameters, in this case nuclear cross sections, will propagate through a model to influence the response or QoI. These are the uncertainties we seek to quantify statistically with Bayesian IUQ. \textit{Experimental uncertainty} in the measured data, in this case the critical eigenvalue, is caused by measurement errors or noise in the data. \textit{Numerical uncertainty} stems from numerical approximation errors such as insufficient spacial or temporal discretization within the model. \textit{Model uncertainty} results from simplifications or inaccuracies in the underlying model that fail to fully capture the physical process. Sometimes model and numerical uncertainties are considered in a combined manner as they may be difficult to separate. \textit{Code uncertainty} (also referred to as \textit{interpolation uncertainty}) is introduced when we use a surrogate model to emulate the responses of a computationally expensive model, for example, with machine learning algorithms, to significantly reduce the computational cost. Ideally any UQ process should account for all sources of uncertainty involved in the modeling process to avoid biased uncertainty estimations.

The formulation of the inverse problem in Eqn.~(\ref{eqn:inv_model}) assumes the model perfectly predicts reality at the true parameter values. This framing does not consider differences between the model $\yM$ and the unknown reality $\yR$. Including model bias $\delta(\bf{x})$ addresses inaccuracies in the model due to improper physics or numerical approximations~\cite{kennedy_2001_bayesian}. 
\begin{equation}
    \yE(\bf{x}) = \yR(\bf{x}) + \bm{\varepsilon} \qquad  \text{where} \qquad \yR(\bf{x}) = \yM(\bf{x},\btheta) + \bm{\delta}(\bf{x})
    \label{eqn:iuq_reality}
\end{equation}
Combining the equations in Eqn.~(\ref{eqn:iuq_reality}) yields the model updating equation as the starting point for IUQ:
\begin{equation}
    \yE(\bf{x}) = \yM(\bf{x},\btheta) + \bm{\delta}(\bf{x}) + \bm{\varepsilon}
    \label{eqn:iuq_updating}
\end{equation}
The \textit{model bias} term (also frequently referred to as \textit{model discrepancy}) is only a function of the design variables $\bu{x}$ and independent from the model parameters. This is mainly because this term exists even if the true parameter values are known.

The strategy of IUQ is to use Markov Chain Monte Carlo (MCMC) to sample directly from the posterior distribution of the parameters. Because the marginal in Eqn.~(\ref{eqn:bayes_theorem}) is not dependent on the parameters $\bm{\theta}$, it will be canceled in MCMC evaluations of sample acceptance criteria and we can rewrite:
\begin{equation}
    \text{p}\pp{\btheta | \yE,\yB} \propto \text{p}\pp{\yE,\yB | \btheta} \cdot \text{p}\pp{\btheta}
    \label{eqn:bayes_propto}
\end{equation}
We replace $\yM$ with $\yB$ in Eqn.~(\ref{eqn:bayes_propto}) because we only have measured data for the \textsc{Experiment} models.

The prior parameter distribution $\text{p}\pp{\btheta}$ represents prior knowledge of the parameters. In this benchmark exercise, the prior is representative of nuclear covariance data and distributed multivariate normal as shown in Table~\ref{tab:prior}. Thus the prior can be written:
\begin{equation}
    \text{p}\pp{\btheta} \propto \exp \brk{ -\frac{1}{2} \pp{\btheta - \bmu_\theta}\T \bm{\Sigma_\theta}^{-1} \brk{\btheta - \bmu_\theta} }
    \label{eqn:iuq_prior}
\end{equation}

In order to form the likelihood $\text{p}\pp{\yE, \yM | \btheta}$, we assume the measurement errors are distributed zero-mean multivariate normal $\bm{\varepsilon} = \pp{\yE - \yM - \bm{\delta}} \sim \cg{N} \pp{0, \bm{\Sigma}}$. This implies our measured data are normally distributed about the model $\yE \sim \cg{N}\pp{\yM + \bm{\delta},\bm{\Sigma}}$ and our likelihood can be shown to be proportional as follows:
\begin{equation}
    \text{p}\pp{\yE, \yB | \btheta} \propto \cdot \frac{1}{\sqrt{|\bm{\Sigma}|}} \exp \brk{ -\frac{1}{2} \pp{\yE-\yB-\bm{\delta}}\T \bm{\Sigma}^{-1} \pp{\yE-\yB-\bm{\delta}} }
    \label{eqn:iuq_likelihood}
\end{equation}
where the total variance is defined as $\bm{\Sigma} = \SigmaE + \bm{\Sigma}_\text{code} + \bm{\Sigma}_\text{bias}$. The $\SigmaE$ term accounts for measurement error, $\bm{\Sigma}_\text{code}$ accounts for the code or interpolation uncertainty introduced by the surrogate model, and $\bm{\Sigma}_\text{bias}$ accounts for the model uncertainty due to model bias. In this exercise the experimental measurements are assumed independent resulting in a diagonal matrix for $\SigmaE$. If a surrogate model is not employed and the full model is used directly by the MCMC sampler, the $\bm{\Sigma}_\text{code}$ term is not needed. Similarly, if the model bias term is not considered then $\bm{\delta}$ and $\bm{\Sigma}_\text{bias}$ are neglected at the risk of overconfidence in the posterior distributions. This is because one is essentially using $\yR(\bf{x}) = \yM(\bf{x},\btheta)$ implying the model predicts the reality exactly.

Mathematically describing the model bias/discrepancy term is always a major challenge for Bayesian calibration. The main cause of this challenge is this term represents the difference between the model and the unknown reality, thus the term itself is never known. One common way is to use the difference between model prediction at nominal parameter values $\bm{\mu}_\theta$ and experimental data, as shown in Eqn.~(\ref{eqn:iuq_discrepancy1}), to serve as the ``training data'' to build a ``surrogate model'' of the model bias.
\begin{equation}
    \bm{\delta} (\bf{x}) = \yE (\bf{x}) - \yB (\bf{x}, \bm{\mu}_\theta)
    \label{eqn:iuq_discrepancy1}
\end{equation}
Because the design space $\bf{x}$ in this benchmark exercise is obscured within each \textsc{Experiment}'s operational conditions, the design space is assumed fixed for each \textsc{Experiment} model and the discrepancy term is simply the difference of the measured data and the model evaluated at the nominal parameters shown in Eqn.~(\ref{eqn:iuq_discrepancy2}).
\begin{equation}
    \bm{\delta} = \yE - \yB \pp{\bm{\mu}_\theta}
    \label{eqn:iuq_discrepancy2}
\end{equation}
It turns out that after dropping the dependence on design variables $\bf{x}$, the model bias term is simply a constant for each \textsc{Experiment}. As a result, the model and numerical uncertainty represented by the covariance of $\bm{\delta}$ in this implementation is simply $\bm{\Sigma}_\text{bias} = \SigmaE$.

For this benchmark exercise, we will compare the IUQ results with and without the model discrepancy term. Together, the model discrepancy term and its covariance will affect both the centering and scaling of the posterior parameter and posterior predictive distributions. It is clear from Eqns.~(\ref{eqn:iuq_updating},~\ref{eqn:iuq_likelihood}) that the centers are shifted to account for bias in the model. The combined shifting and scaling of the posterior distributions is seen in Section~\ref{subsection:Method-Comparison} when comparing IUQ results with and without the model discrepancy term. This comparison also demonstrates the increased variance in the Bayesian inference process does not necessarily result in an increased posterior predictive distribution variance. When models exhibit nonlinear behaviors, shifting the posterior predictive distribution may place it in a region of lower variance, despite having similar or larger variance in the posterior parameter distribution.

With the prior, likelihood, and model discrepancy defined, we have the elements to complete the IUQ process. For this project, we generally follow the improved modular Bayesian approach outlined in~\cite{wu_2018_inverse1, wu_2018_inverse2}. We allocate our data, build a Gaussian Process surrogate for the models, perform MCMC to get the posterior parameter distributions, then perform FUQ using these posterior parameter distributions to obtain the posterior predictive distributions of the QoIs.

\section{Results \& Discussion}
\label{section:Results}

The prior model behaviors are explored before presenting analysis of the posterior adjustments and comparisons of the cases and methodologies' performance. In order to extract as much information as possible from this performance benchmark exercise, each \textsc{Experiment} that is not used in the respective case will be treated as an \textsc{Application} in that case. For instance: in \textsc{Case 1}, only measured data from \texttt{Albert} is used for inference thus \texttt{Bohr}, \texttt{Chadwick}, and \texttt{Dyson} will be treated as \textsc{Applications} in Case 1. This approach will be used for every case, as it allows comparisons of posterior predictions with experimentally measured data.

\subsection{Prior Model Behavior}
\label{subsection:Prior-Model-Behavior}

\begin{figure}[!htb]
    \centering
    \includegraphics[width=0.9\linewidth]{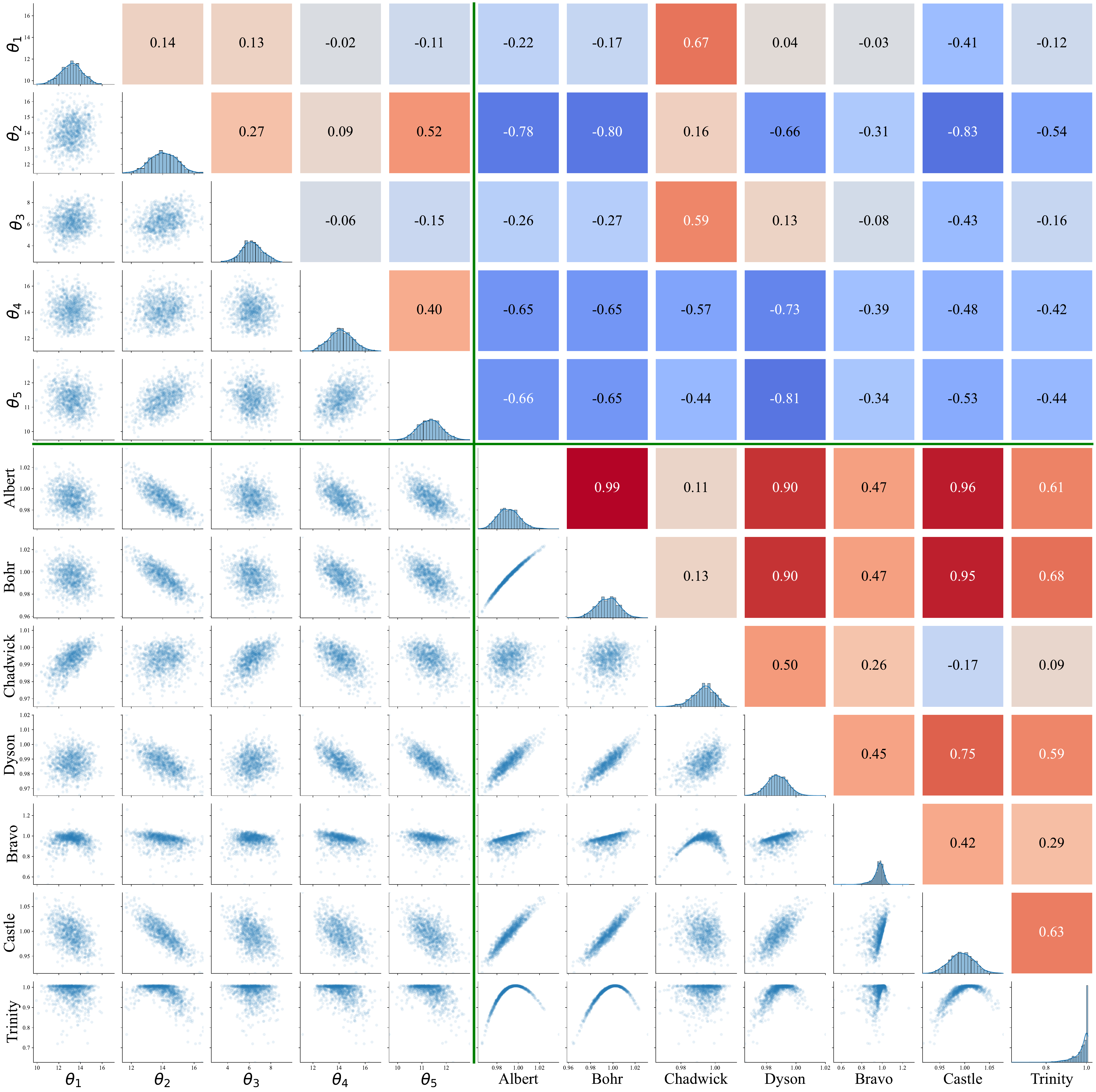}
    \caption{Pairwise plot of prior parameter and corresponding model response distributions with correlation coefficients in upper right.}
    \label{fig:pairwise_all_prior}
\end{figure}

We begin the analysis by examining the model behavior over the domain of the prior parameter distribution. This allows validation of the assumptions made by each of our methods. Fig.~\ref{fig:pairwise_all_prior} shows pairwise plots for parameter samples drawn from the prior distribution and their corresponding model responses with scatterplots in the lower triangle and (Pearson) correlation coefficients in the upper triangle. The green lines delineate parameters from responses, dividing the pairwise plots into four distinct boxes. This division shows the prior parameter distributions in the top left box and the prior model response distributions in the lower right box. The upper right box shows the correlation coefficient of the models to parameters. The lower left box shows scatter plots of the model responses with the respective parameters. As demonstrated in~\cite{saltelli_2008_global}, both the lower left and upper right boxes provide a very quick indication of model sensitivity to the parameters. While scatter plots are very effective indications of sensitivity except in very special cases, the correlation coefficient is only truly informative in near linear models.

These pairwise plots reveal much about the model behaviors. The upper left box is simply a confirmation of multivariate normal prior distribution of the parameters $\btheta$. Examining the model response distributions in the lower right box reveals two important observations about \textsc{Applications} \texttt{Bravo} and \texttt{Trinity}: (1) they are not normally distributed, and (2) they demonstrate nonlinear behaviors over the parameter domain in the scatterplots. Both observations conflict with the underlying linearity assumption of GLLS, indicating that the GLLS results may be questionable for these nonlinear \textsc{Applications}.

Further examining the scatterplots in the lower left box shows nonlinear relationships for \texttt{Trinity} in $\theta_2$ and $\theta_5$, which correspond to strong linear relationships in \texttt{Albert}, \texttt{Bohr}, \texttt{Dyson}, and \texttt{Castle} for those parameters. This indicates the underlying cause of the strong nonlinear dependencies between these models shown in the lower right box, though the model behavior over the entire parameter domain must be considered. A similar observation can be made for \texttt{Bravo} and \texttt{Chadwick} in $\theta_1$. The scatterplots in the lower left also indicate very similar linear model responses and model sensitivity profiles over the parameter domain for \texttt{Albert}, \texttt{Bohr}, \texttt{Dyson}, and \texttt{Castle}. However, the corresponding scatterplots for \texttt{Chadwick} show very different sensitivity profiles over the parameter domain. While \texttt{Chadwick} still demonstrates linear behaviors, it shows strong dependence on $\theta_1$ and $\theta_3$ instead of $\theta_2$, $\theta_4$, and $\theta_5$. Furthermore, \texttt{Chadwick}'s dependence on the first three parameters is slightly inverse of \texttt{Albert}, \texttt{Bohr}, \texttt{Dyson}, and \texttt{Castle}. As a result, \texttt{Chadwick} has low correlation with the other models and anti-correlation with \texttt{Castle}.

Comparing the correlation of \textsc{Experiment} and \textsc{Applications} shown in the upper triangle of the lower right box in Fig.~\ref{fig:pairwise_all_prior} with the $c_k$ values provided in Table~\ref{tab:correlation} show general agreement for \texttt{Castle}, which we observed to have linear behaviors. However, they show disagreement for \texttt{Bravo} and \texttt{Trinity}, which we observed to have nonlinear behaviors. Similarly, when we compare the model standard deviations provided in Table~\ref{tab:responses} with those computed from these prior samples, we again get general agreement for models with linear behaviors and strong disagreement for those models with nonlinear behaviors. A comparison of these model standard deviations is summarized in Table~\ref{tab:std_comparison}. These discrepancies are easily attributed to the assumption that the model is linear around the the gradient evaluated at the prior parameter mean and subsequent linear approximation of the model covariances and correlations in Eqn.~(\ref{eqn:sandwich_rule}).

\begin{table}[!htb]
    \centering
    \caption{Comparison of model standard deviation from linear approximation and prior samples.}
    \begin{tabular}{l|c c c c c c c|}
        & \multicolumn{7}{c|}{\textbf{Model Standard Deviation}} \\
        & \texttt{Albert} & \texttt{Bohr} & \texttt{Chadwick} & \texttt{Dyson} & \texttt{Bravo} & \texttt{Castle} & \texttt{Trinity} \\ \hline
        \textbf{Linear Approx}     &  0.01002 & 0.01002 &  0.00623 &  0.00701 &  0.02900 & 0.02494 & 0.03022  \\
        \textbf{Prior sample}  &  0.01004 & 0.01001 &  0.00632 &  0.00702 &  0.05791 & 0.02490 & 0.04379  \\ \hline
        \textbf{Difference} & -0.00002 & 0.00001 & -0.00009 & -0.00001 & -0.02891 & 0.00004 & -0.0136  \\ \hline
    \end{tabular}
    \label{tab:std_comparison}
\end{table}

These observations support the validity of GLLS for linear models while demonstrating the need for more robust methods for nonlinear models. They also demonstrate the insufficiency of model correlation $c_k$ as a metric of similarity for models exhibiting nonlinear behaviors. Correlation indicates linear dependence but is a poor indicator of nonlinear dependence. Any metric or statistic used to assess model dependence needs to capture nonlinear dependence, if one is required at all. Model dependence is strongly influenced by model sensitivities to the input parameters, thus model sensitivities might be a better indication of \textsc{Experiment} relevance to a data adjustment. Some intuition for this is provided in Section~\ref{subsection:IUQ-Observations}.

\subsection{Method Comparison for Data Assimilation}
\label{subsection:Method-Comparison}

Each method's performance is compared for the posterior parameters and posterior predictions. Then the posterior predictive distributions of GLLS and MOCABA are compared with actual model evaluations at parameters sampled from that method's posterior parameters distribution. IUQ uses the actual model evaluations for posterior parameter samples to produce the empirical posterior distributions, making its posterior predictive distributions nearly identical to the computed model evaluations.

The posterior parameter distributions for each case and method are shown in Table~\ref{tab:posterior_parameters}. While this table does not capture the full context of parameter relationships contained in their covariance matrices, it serves as a quick comparison of the data adjustments. The posterior parameter values are quite comparable across methods with a slight deviation for IUQ when the model discrepancy is included. This is expected given all 4 of the \textsc{Experiment} models exhibit linear behavior over the domain of the prior parameter distributions. In GLLS and MOCABA, the posterior parameter distributions computed in Eqns.~(\ref{eqn:glls_param},~\ref{eqn:mocaba_param}) are only dependent on the linear \textsc{Experiment} models and not on the nonlinear \textsc{Application} models. An interesting exercise not explored in this benchmark exercise would compare methods when there are nonlinear behaviors in the \textsc{Experiment} models. The posterior parameter distributions from IUQ with model discrepancy $\delta$ are shifted due to the recentering resulting from consideration of model bias.

\begin{table}[!htb]
    \centering
    \caption{Prior and posterior parameter mean values and standard deviations for each case and method.}
    \begin{tabular}{l l|c c|c c|c c|c c|c c|}
        & & \multicolumn{2}{c|}{$\theta_1$} & \multicolumn{2}{c|}{$\theta_2$} & \multicolumn{2}{c|}{$\theta_3$} & \multicolumn{2}{c|}{$\theta_4$} & \multicolumn{2}{c|}{$\theta_5$} \\
        & & mean & std & mean & std & mean & std & mean & std & mean & std \\ \hline
        \multicolumn{2}{c|}{\textbf{Prior}} & 13.14 & 1.00 & 14.06 & 0.88 & 6.27 & 0.97 & 14.13 & 0.87 & 11.32 & 0.49 \\ \hline
        \multirow{4}{*}{\textbf{Case 1}}  
          & GLLS             & 12.98 & 0.97 & 13.59 & 0.54 & 6.12 & 0.94 & 13.76 & 0.67 & 11.10 & 0.36 \\
          & MOCABA           & 12.99 & 0.97 & 13.60 & 0.54 & 6.11 & 0.94 & 13.77 & 0.68 & 11.11 & 0.36 \\
          & IUQ w/o $\delta$ & 13.00 & 0.88 & 13.61 & 0.50 & 6.12 & 0.89 & 13.80 & 0.61 & 11.10 & 0.34 \\
          & IUQ              & 13.20 & 0.92 & 14.08 & 0.48 & 6.25 & 0.86 & 14.13 & 0.58 & 11.35 & 0.33 \\ \hline
        \multirow{4}{*}{\textbf{Case 2}}  
          & GLLS             & 12.93 & 0.92 & 13.61 & 0.52 & 6.14 & 0.93 & 13.76 & 0.67 & 11.10 & 0.36 \\
          & MOCABA           & 13.01 & 0.96 & 13.57 & 0.51 & 6.10 & 0.94 & 13.76 & 0.68 & 11.10 & 0.36 \\
          & IUQ w/o $\delta$ & 13.08 & 0.86 & 13.61 & 0.49 & 6.15 & 0.85 & 13.73 & 0.62 & 11.12 & 0.33 \\
          & IUQ              & 13.18 & 0.88 & 14.07 & 0.47 & 6.28 & 0.86 & 14.12 & 0.60 & 11.33 & 0.35 \\ \hline
        \multirow{4}{*}{\textbf{Case 3}}  
          & GLLS             & 13.49 & 0.69 & 13.75 & 0.49 & 6.54 & 0.76 & 13.43 & 0.50 & 10.97 & 0.31 \\
          & MOCABA           & 13.48 & 0.73 & 13.72 & 0.50 & 6.50 & 0.79 & 13.45 & 0.53 & 10.97 & 0.31 \\
          & IUQ w/o $\delta$ & 13.55 & 0.63 & 13.80 & 0.45 & 6.55 & 0.68 & 13.42 & 0.46 & 10.97 & 0.29 \\
          & IUQ              & 13.12 & 0.64 & 14.09 & 0.48 & 6.29 & 0.69 & 14.15 & 0.50 & 11.32 & 0.30 \\ \hline
        \multirow{4}{*}{\textbf{Case 4}}  
          & GLLS             & 13.30 & 0.86 & 13.65 & 0.53 & 6.64 & 0.59 & 13.56 & 0.61 & 10.94 & 0.28 \\
          & MOCABA           & 13.21 & 0.91 & 13.59 & 0.54 & 6.45 & 0.78 & 13.61 & 0.63 & 10.98 & 0.30 \\
          & IUQ w/o $\delta$ & 13.34 & 0.78 & 13.69 & 0.50 & 6.67 & 0.53 & 13.56 & 0.57 & 10.95 & 0.26 \\
          & IUQ              & 13.22 & 0.76 & 14.05 & 0.49 & 6.24 & 0.58 & 14.17 & 0.56 & 11.31 & 0.26 \\ \hline
        \multirow{4}{*}{\textbf{Case 5}}  
          & GLLS             & 13.39 & 0.60 & 13.75 & 0.43 & 6.67 & 0.53 & 13.45 & 0.47 & 10.94 & 0.28 \\
          & MOCABA           & 13.49 & 0.56 & 13.75 & 0.45 & 6.50 & 0.71 & 13.42 & 0.50 & 10.97 & 0.30 \\
          & IUQ w/o $\delta$ & 13.55 & 0.55 & 13.74 & 0.40 & 6.60 & 0.46 & 13.47 & 0.44 & 10.95 & 0.25 \\
          & IUQ              & 13.17 & 0.60 & 14.05 & 0.42 & 6.26 & 0.55 & 14.15 & 0.46 & 11.33 & 0.26 \\ \hline
    \end{tabular}
    \label{tab:posterior_parameters}
\end{table}

\begin{figure}[!htb]
    \centering
    \begin{subfigure}{0.49\linewidth}
        \centering
        \includegraphics[width=\linewidth]{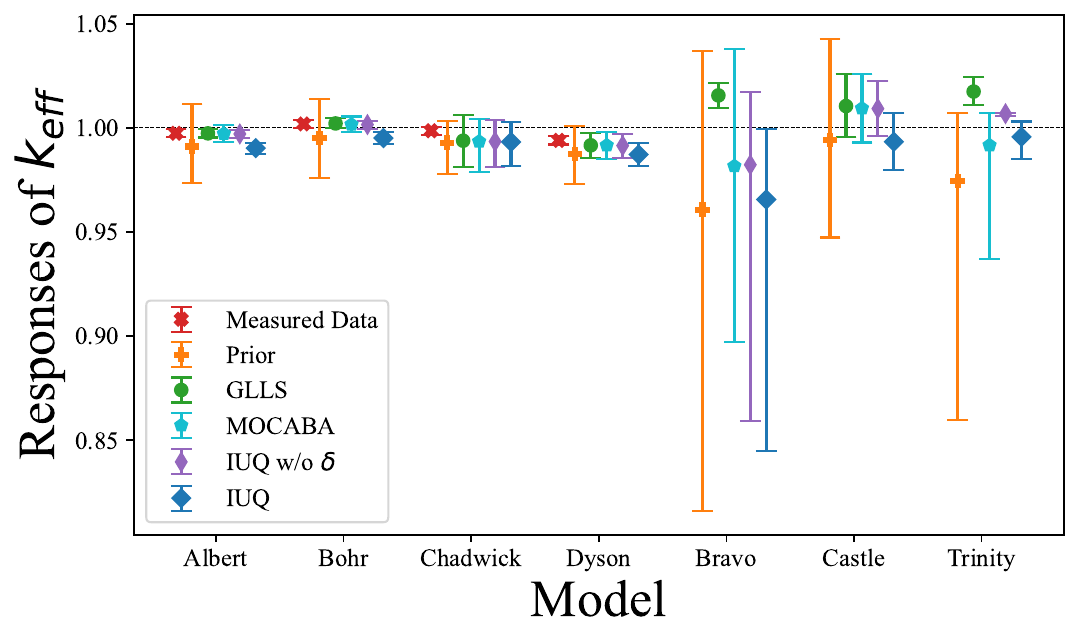}
        \caption{\centering Case 1: \texttt{Albert}}
    \end{subfigure}
    \begin{subfigure}{0.49\linewidth}
        \centering
        \includegraphics[width=\linewidth]{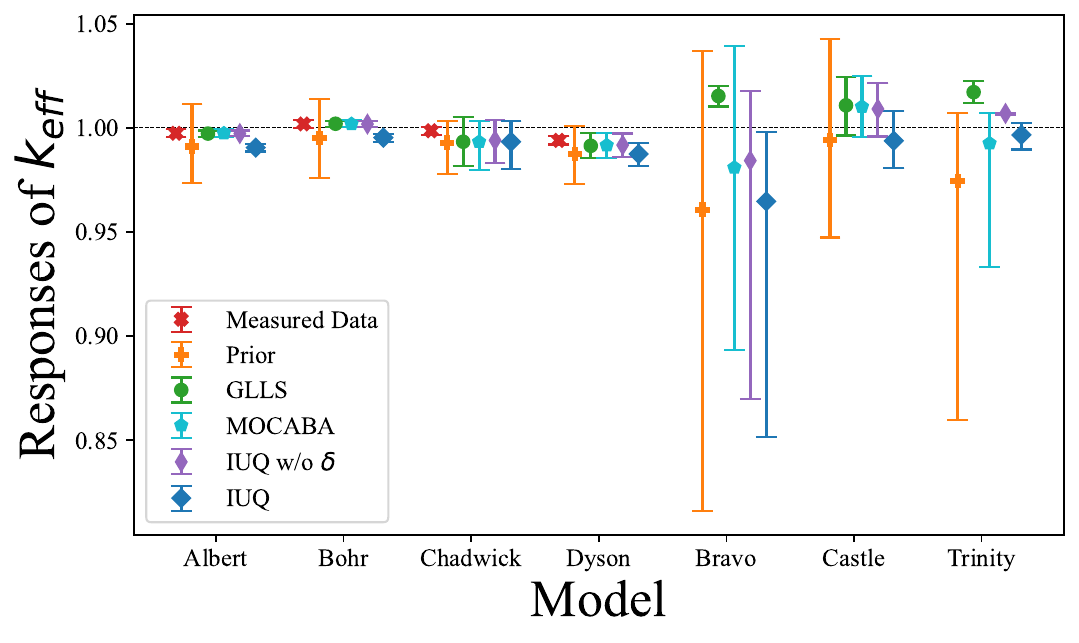}
        \caption{\centering Case 2: \texttt{Albert} + \texttt{Bohr}}
    \end{subfigure}
    \begin{subfigure}{0.49\linewidth}
        \centering
        \includegraphics[width=\linewidth]{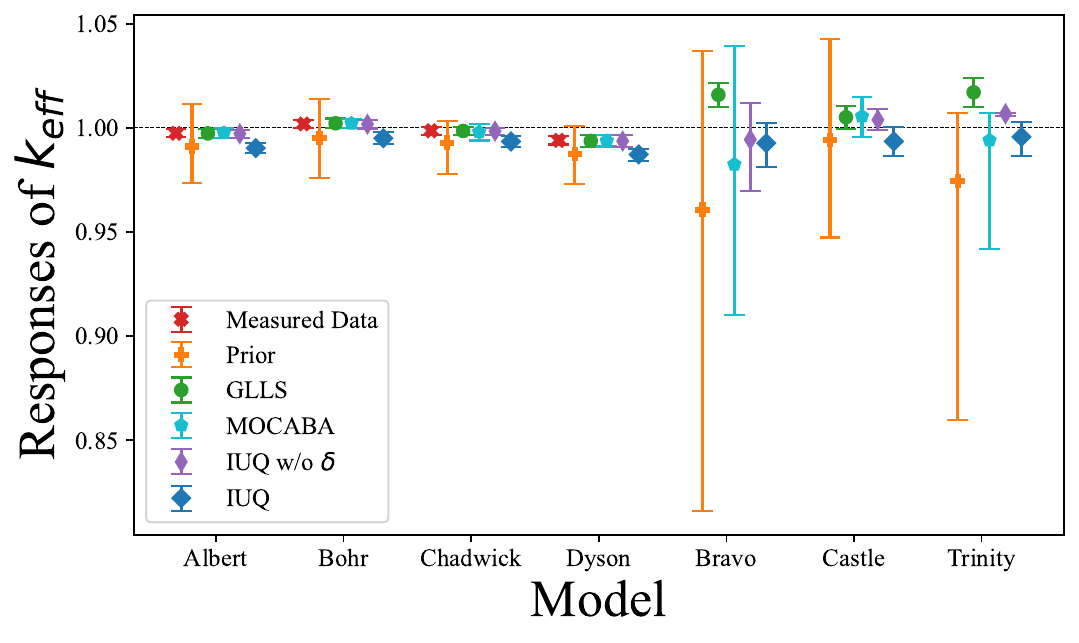}
        \caption{\centering  Case 3: \texttt{Albert} + \texttt{Chadwick}}
    \end{subfigure}
    \begin{subfigure}{0.49\linewidth}
        \centering
        \includegraphics[width=\linewidth]{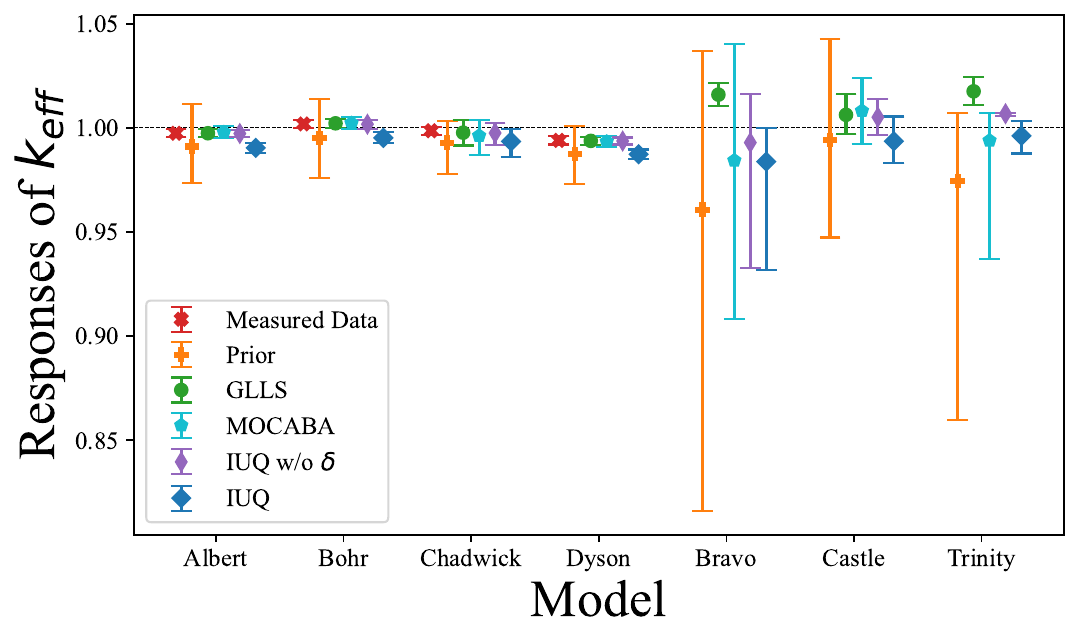}
        \caption{\centering Case 4: \texttt{Albert} + \texttt{Dyson}}
    \end{subfigure}
    \begin{subfigure}{0.49\linewidth}
        \centering
        \includegraphics[width=\linewidth]{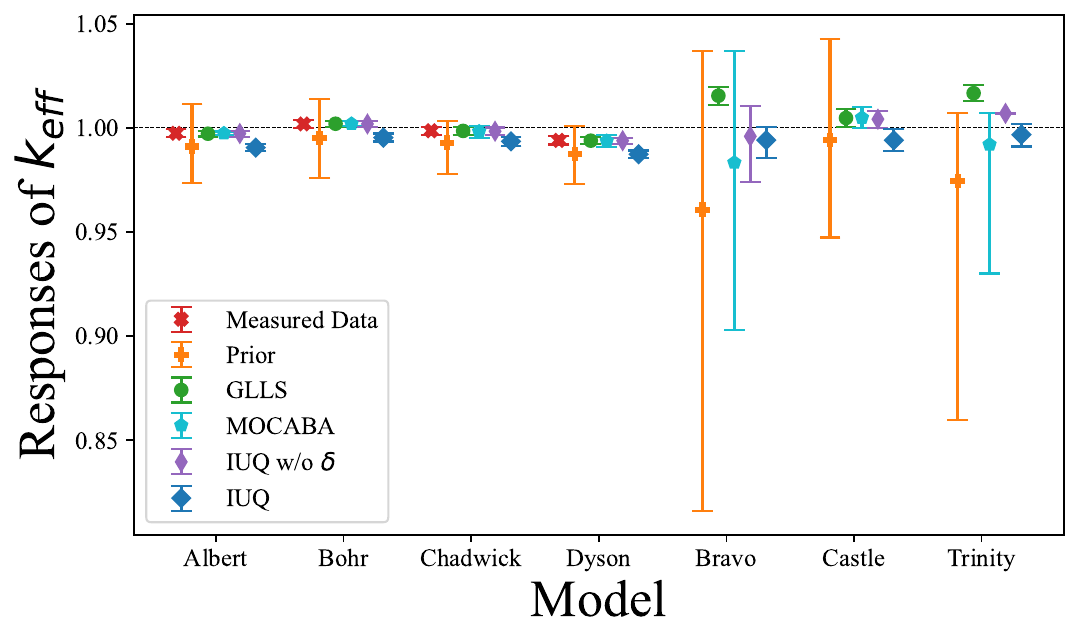}
        \caption{\centering Case 5: All}
    \end{subfigure}
    \caption{Comparison across methods of posterior predictions with 95\% credible intervals.}
    \label{fig:compare_predictions}
\end{figure}

The posterior predictions for each case and method are shown in Table~\ref{tab:posterior_predictions} with associated plots shown in Fig.~\ref{fig:compare_predictions}. Because \textsc{Experiments} not used in that case's data adjustment are treated as \textsc{Applications}, Cases 1-4 provide a comparison of posterior predictions to measured data of \texttt{Bohr}, \texttt{Chadwick}, or \texttt{Dyson}. In these instances, posterior prediction credible intervals which cover the measured data of these \textsc{Experiments} are given some validation. Strong agreement between GLLS, MOCABA, and Bayesian IUQ is observed in posterior predictions for \textsc{Applications} exhibiting linear behaviors. However, significant deviations are observed in posterior predictions of the two \textsc{Applications} with nonlinear behaviors (\texttt{Bravo} and \texttt{Trinity}). This serves to reaffirm the validity of GLLS for linear \textsc{Applications} while highlighting the failure of its underlying assumptions when applied to nonlinear \textsc{Applications}.

\begin{table}[!htb]
    \centering
    \caption{\textsc{Application} posterior predictions with 95\% credible intervals for each case and method.}
    \begin{tabular}{l l|c c c|c c c|c c c|}
        & & \multicolumn{3}{c|}{\texttt{Bravo}} & \multicolumn{3}{c|}{\texttt{Castle}} & \multicolumn{3}{c|}{\texttt{Trinity}} \\
        & & mean & lower & upper & mean & lower & upper & mean & lower & upper \\ \hline
        \multicolumn{2}{c|}{\textbf{Prior}} & 1.0000 & 0.9999 & 1.0001 & 1.0000 & 0.9999 & 1.0001 & 1.0000 & 0.9999 & 1.0001 \\ \hline
        \multirow{4}{*}{\textbf{Case 1}}  
          & GLLS             & 1.0155 & 1.0096 & 1.0214 & 1.0103 & 0.9958 & 1.0249 & 1.0174 & 1.0108 & 1.0241 \\
          & MOCABA           & 0.9917 & 0.9851 & 0.9980 & 0.9817 & 0.8972 & 1.0377 & 1.0094 & 0.9930 & 1.0258 \\
          & IUQ w/o $\delta$ & 0.9915 & 0.9856 & 0.9972 & 0.9823 & 0.8594 & 1.0172 & 1.0093 & 0.9961 & 1.0228 \\
          & IUQ              & 0.9873 & 0.9818 & 0.9928 & 0.9656 & 0.8451 & 0.9997 & 0.9934 & 0.9799 & 1.0074 \\ \hline
        \multirow{4}{*}{\textbf{Case 2}}  
          & GLLS             & 1.0152 & 1.0105 & 1.0199 & 1.0102 & 0.9957 & 1.0247 & 1.0171 & 1.0121 & 1.0221 \\
          & MOCABA           & 0.9917 & 0.9857 & 0.9978 & 0.9810 & 0.8934 & 1.0393 & 1.0102 & 0.9956 & 1.0248 \\
          & IUQ w/o $\delta$ & 0.9918 & 0.9863 & 0.9974 & 0.9843 & 0.8700 & 1.0176 & 1.0091 & 0.9959 & 1.0214 \\
          & IUQ              & 0.9875 & 0.9819 & 0.9927 & 0.9648 & 0.8517 & 0.9982 & 0.9938 & 0.9807 & 1.0082 \\ \hline
        \multirow{4}{*}{\textbf{Case 3}}  
          & GLLS             & 1.0159 & 1.0101 & 1.0217 & 1.0052 & 0.9999 & 1.0104 & 1.0171 & 1.0105 & 1.0237 \\
          & MOCABA           & 0.9938 & 0.9911 & 0.9967 & 0.9825 & 0.9101 & 1.0392 & 1.0056 & 0.9958 & 1.0149 \\
          & IUQ w/o $\delta$ & 0.9937 & 0.9908 & 0.9965 & 0.9944 & 0.9699 & 1.0120 & 1.0039 & 0.9990 & 1.0089 \\
          & IUQ              & 0.9873 & 0.9843 & 0.9901 & 0.9927 & 0.9811 & 1.0025 & 0.9935 & 0.9867 & 1.0005 \\ \hline
        \multirow{4}{*}{\textbf{Case 4}}  
          & GLLS             & 1.0161 & 1.0105 & 1.0217 & 1.0064 & 0.9966 & 1.0162 & 1.0177 & 1.0111 & 1.0243 \\
          & MOCABA           & 0.9934 & 0.9907 & 0.9963 & 0.9844 & 0.9084 & 1.0403 & 1.0082 & 0.9924 & 1.0239 \\
          & IUQ w/o $\delta$ & 0.9937 & 0.9921 & 0.9954 & 0.9930 & 0.9326 & 1.0162 & 1.0052 & 0.9967 & 1.0141 \\
          & IUQ              & 0.9874 & 0.9851 & 0.9897 & 0.9838 & 0.9318 & 1.0000 & 0.9936 & 0.9832 & 1.0055 \\ \hline
        \multirow{4}{*}{\textbf{Case 5}}  
          & GLLS             & 1.0156 & 1.0111 & 1.0201 & 1.0049 & 1.0007 & 1.0092 & 1.0168 & 1.0127 & 1.0210 \\
          & MOCABA           & 0.9938 & 0.9909 & 0.9967 & 0.9834 & 0.9028 & 1.0368 & 1.0049 & 1.0002 & 1.0100 \\
          & IUQ w/o $\delta$ & 0.9939 & 0.9924 & 0.9952 & 0.9960 & 0.9740 & 1.0107 & 1.0042 & 1.0002 & 1.0082 \\
          & IUQ              & 0.9874 & 0.9855 & 0.9892 & 0.9942 & 0.9858 & 1.0004 & 0.9941 & 0.9889 & 0.9993 \\ \hline
    \end{tabular}
    \label{tab:posterior_predictions}
\end{table}

When comparing the adjustments by case, strong similarities are observed for Cases 1, 2, and 4 with regard to both the posterior parameter distributions and the posterior predictions. However, significant changes to the data adjustments are observed when \texttt{Chadwick} is included for Cases 3 and 5. Despite Table~\ref{tab:correlation} showing a low or even negative correlation coefficient between \texttt{Chadwick} and the \textsc{Applications}, it appears to significantly inform the posterior distributions. This observation is discussed in more detail in Section~\ref{subsection:IUQ-Observations}.

\begin{figure}[!htb]
    \centering
    \begin{subfigure}{0.32\linewidth}
        \centering
        \includegraphics[width=\linewidth]{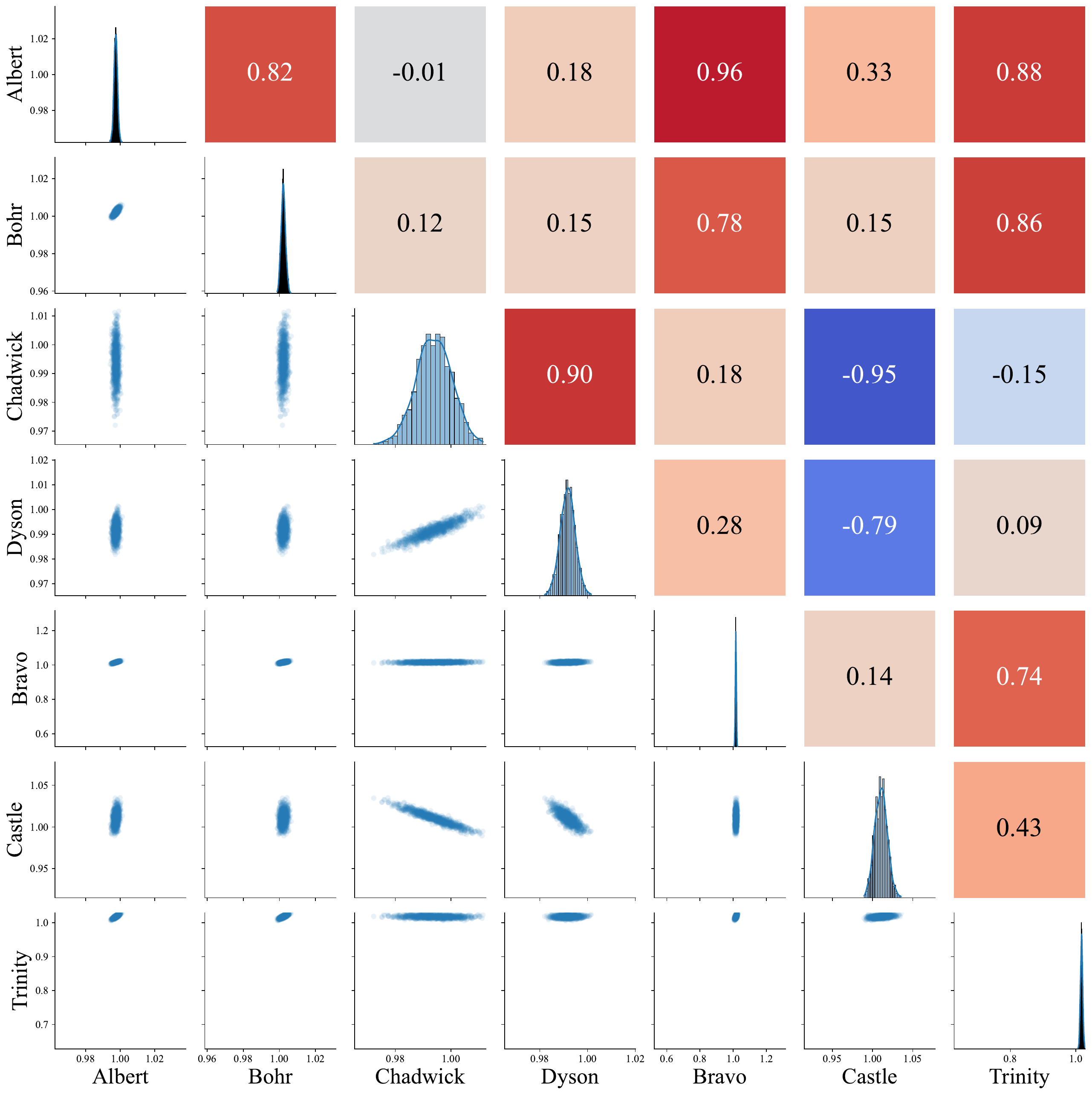}
        \caption{\centering Case 1: Predicted}
    \end{subfigure}
    \begin{subfigure}{0.32\linewidth}
        \centering
        \includegraphics[width=\linewidth]{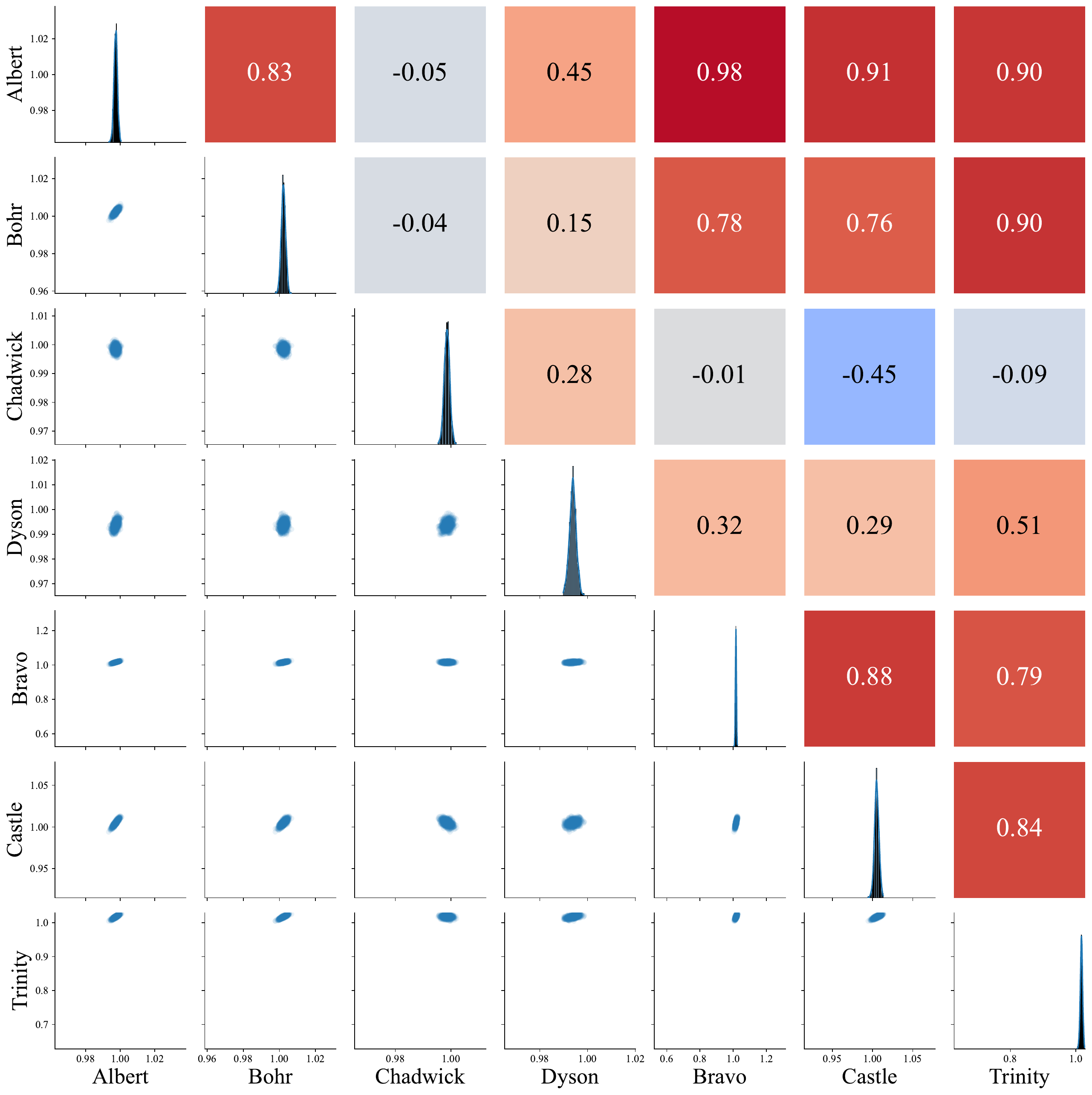}
        \caption{\centering Case 3: Predicted}
    \end{subfigure}
    \begin{subfigure}{0.32\linewidth}
        \centering
        \includegraphics[width=\linewidth]{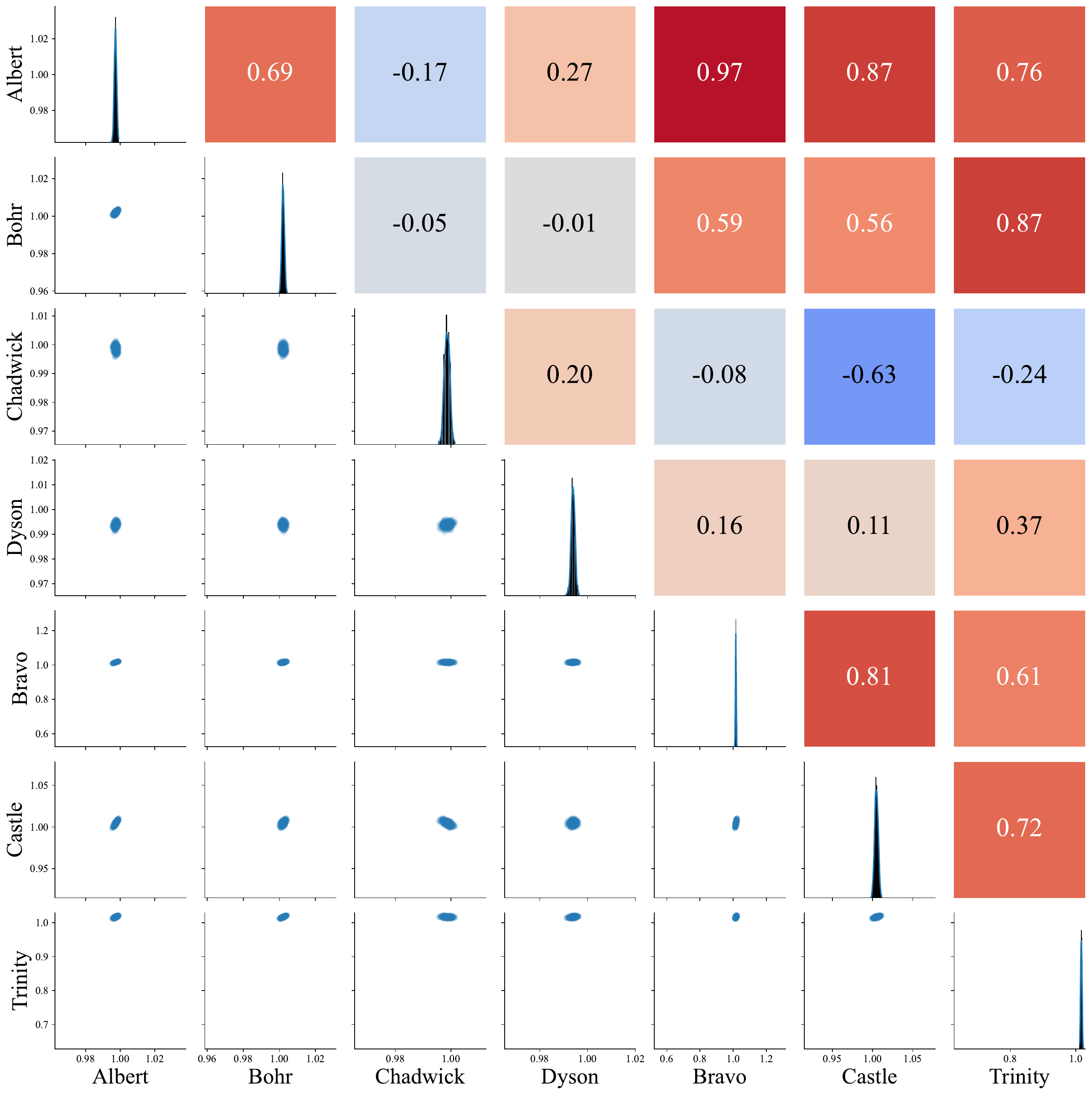}
        \caption{\centering Case 5: Predicted}
    \end{subfigure}
    \begin{subfigure}{0.32\linewidth}
        \centering
        \includegraphics[width=\linewidth]{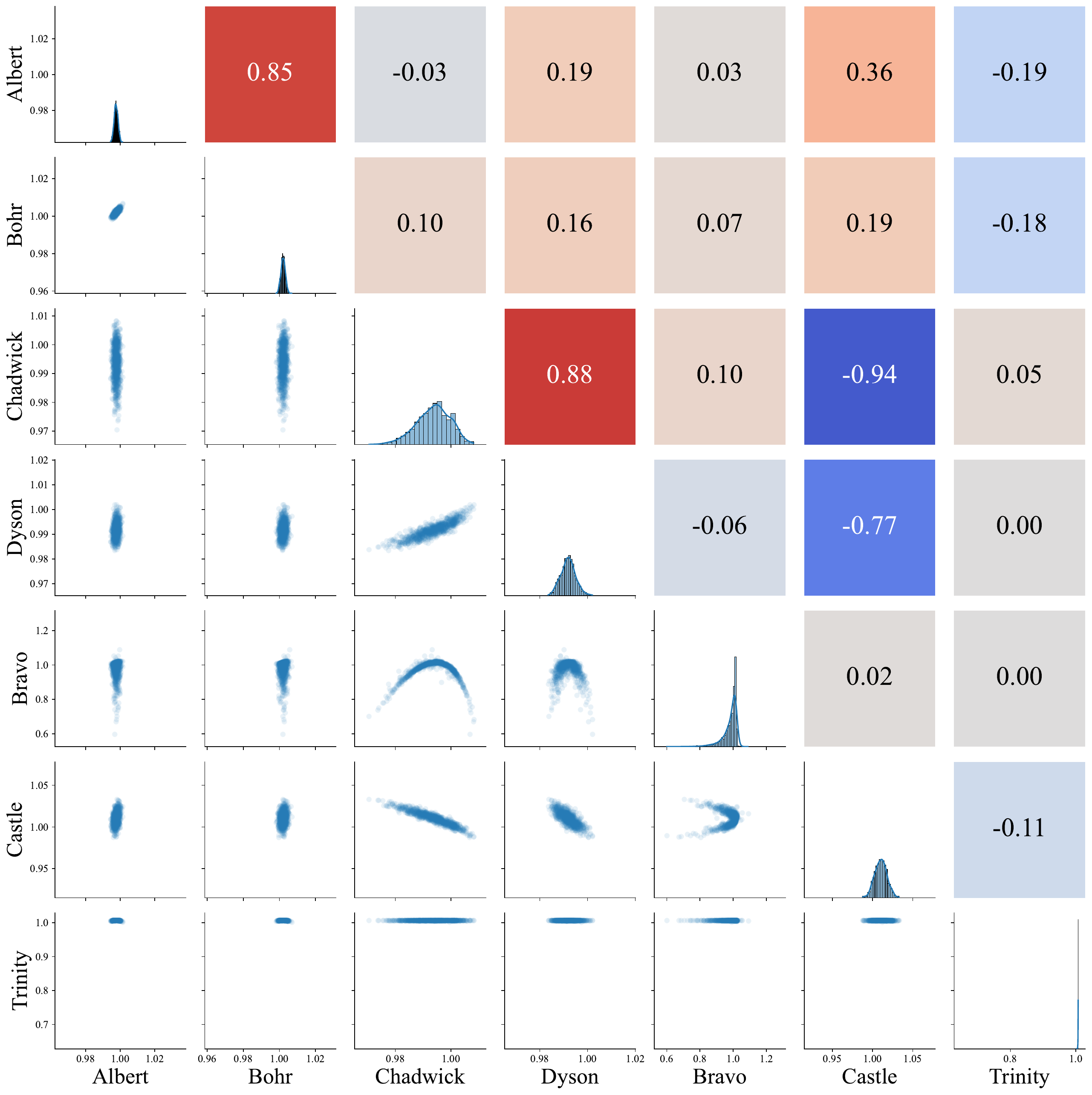}
        \caption{\centering Case 1: Computed}
    \end{subfigure}
    \begin{subfigure}{0.32\linewidth}
        \centering
        \includegraphics[width=\linewidth]{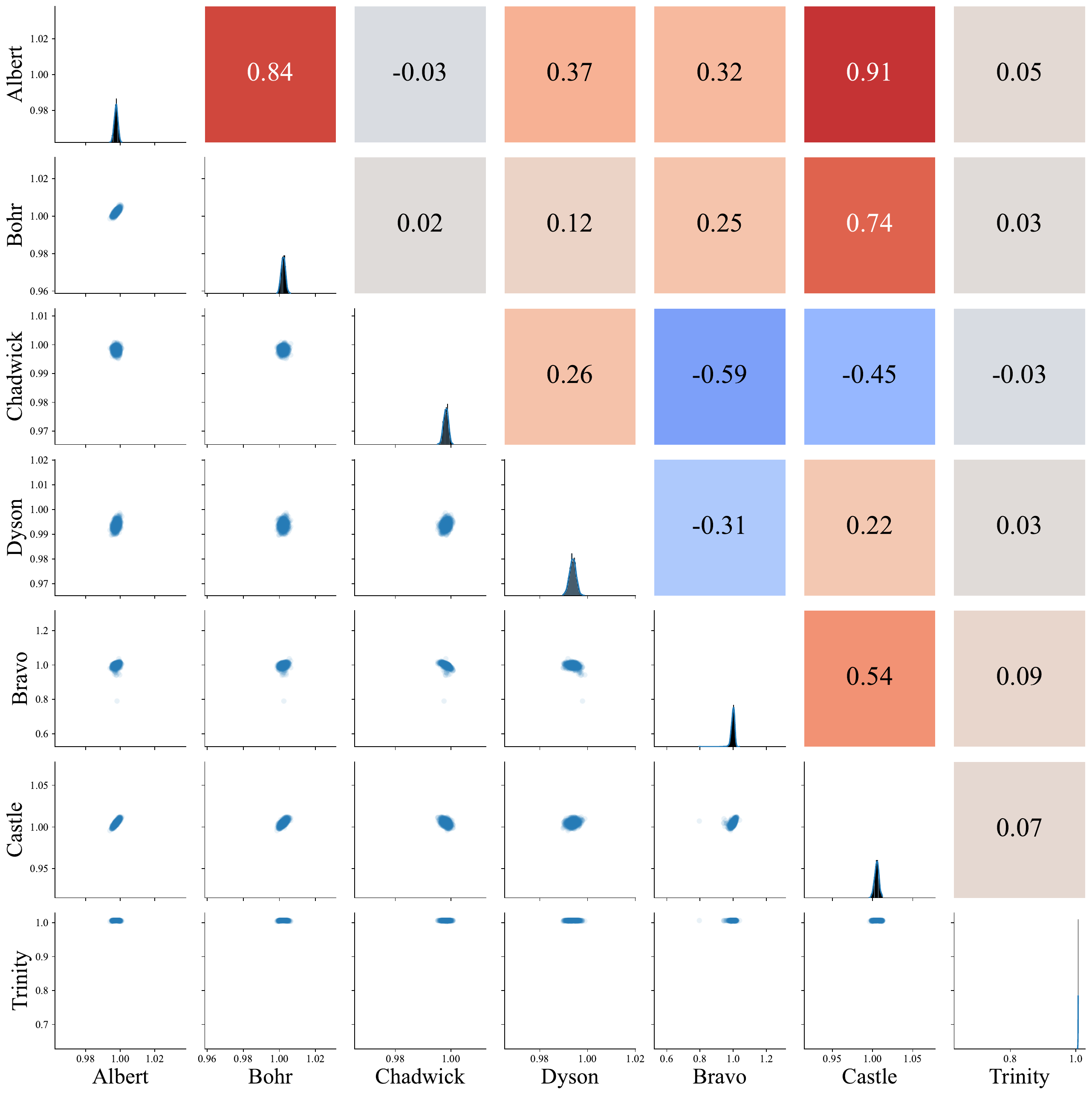}
        \caption{\centering Case 3: Computed}
    \end{subfigure}
    \begin{subfigure}{0.329\linewidth}
        \centering
        \includegraphics[width=\linewidth]{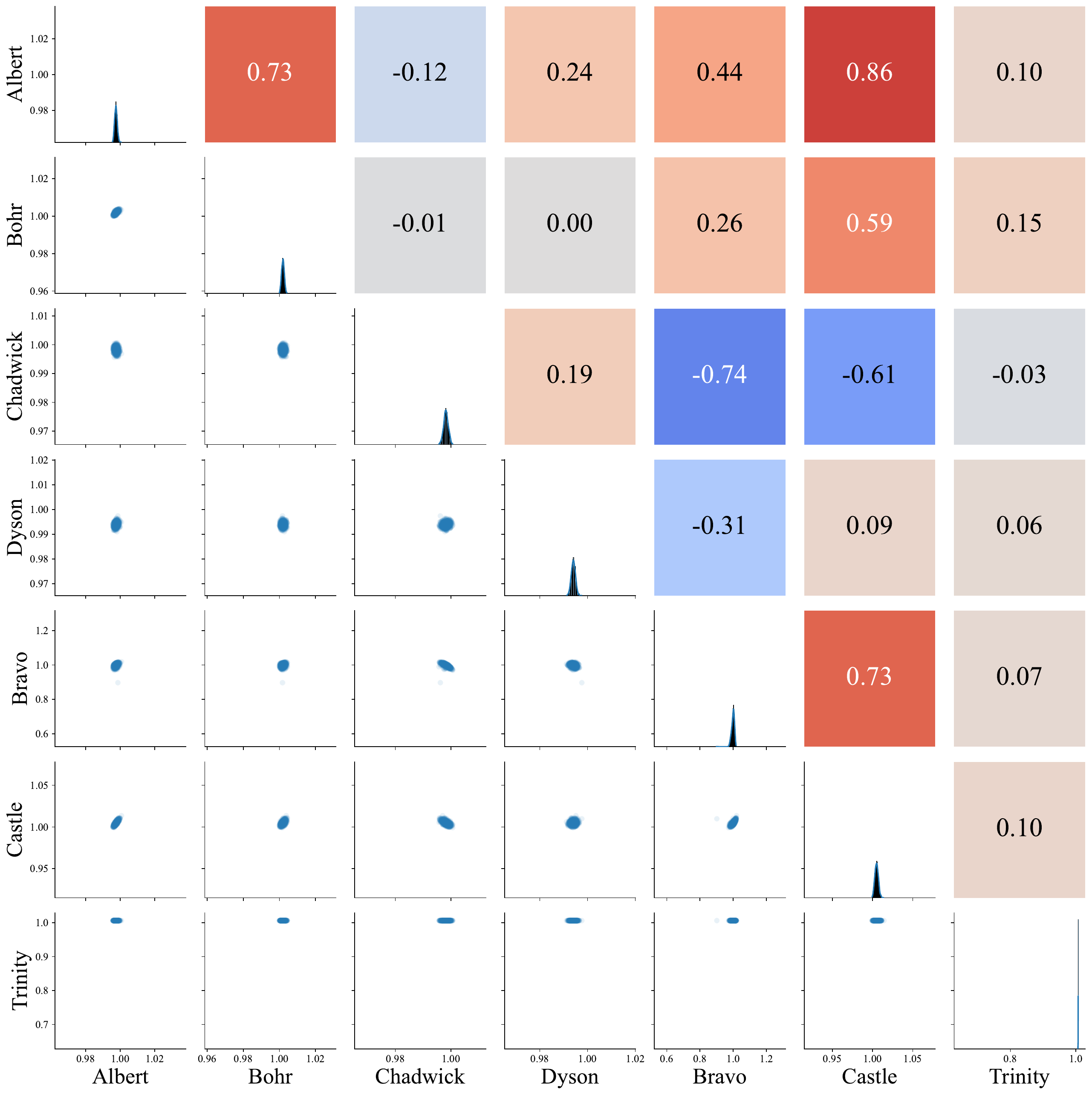}
        \caption{\centering Case 5: Computed}
    \end{subfigure}
    \caption{Comparison of predicted and computed GLLS posterior distributions.}
    \label{fig:compare_glls}
\end{figure}

Pairwise plots for samples from the posterior predictive distributions and computed predictions at posterior parameter samples are shown for GLLS in Fig.~\ref{fig:compare_glls} with plots occurring on the same scale for ease of comparison. We expect reliable posterior predictive distributions to match the distributions of computed predictions. These plots highlight how well, or poorly, the method's posterior prediction matches the distribution of responses computed at samples from the posterior parameter distribution. Because Cases 1, 2, and 4 are quite similar, only Cases 1, 3, and 5 are shown for comparison here. Again, GLLS shows agreement between posterior predictions and computed responses for linear \textsc{Applications}. For Case 1 the linear applications are \texttt{Bohr}, \texttt{Chadwick}, \texttt{Dyson}, and \texttt{Castle}. For Case 3 the linear applications are \texttt{Bohr}, \texttt{Dyson}, and \texttt{Castle}. For Case 5 the linear application is \texttt{Castle}. However, GLLS fails to capture non-normal posterior predictive distributions in the nonlinear \textsc{Applications} \texttt{Bravo} and \texttt{Trinity}. The linear transformations inherent to GLLS only provide information on mean and covariance resulting in normally distributed posterior predictive distributions without regard for skewness or kurtosis in contrast with the computed response distributions.

\begin{figure}[!htb]
    \centering
    \begin{subfigure}{0.32\linewidth}
        \centering
        \includegraphics[width=\linewidth]{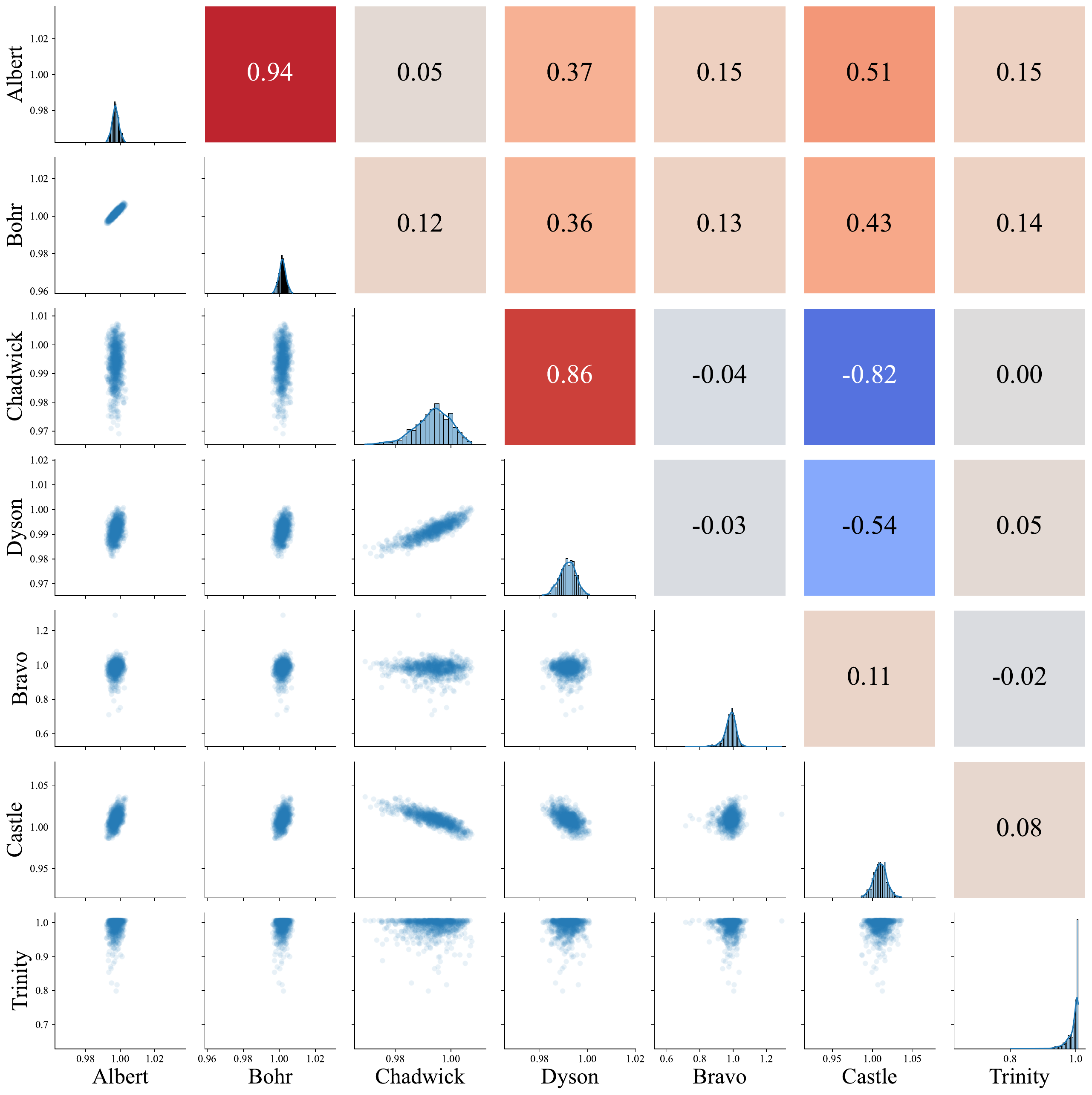}
        \caption{\centering Case 1: Predicted}
    \end{subfigure}
    \begin{subfigure}{0.32\linewidth}
        \centering
        \includegraphics[width=\linewidth]{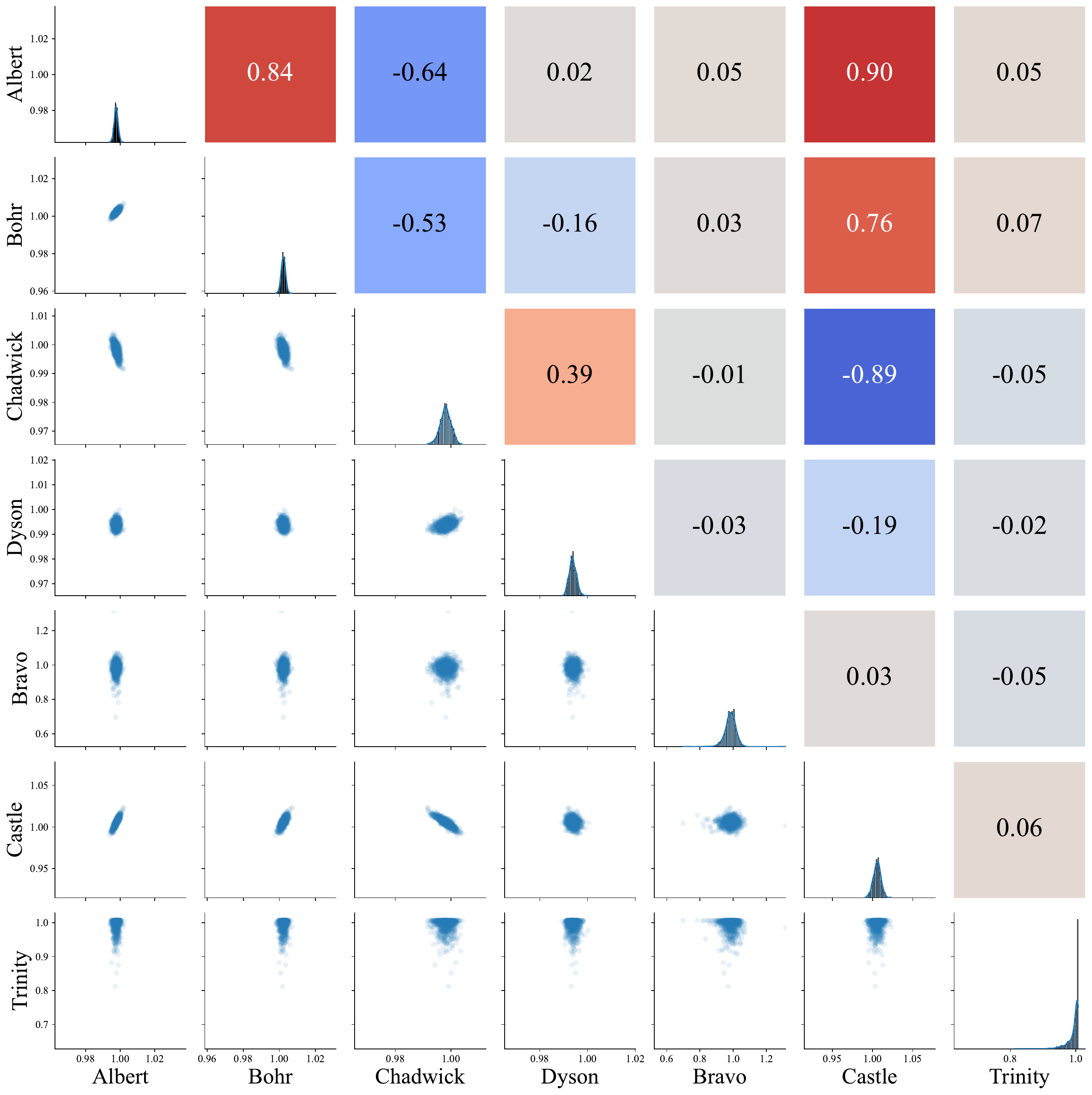}
        \caption{\centering Case 3: Predicted}
    \end{subfigure}
    \begin{subfigure}{0.32\linewidth}
        \centering
        \includegraphics[width=\linewidth]{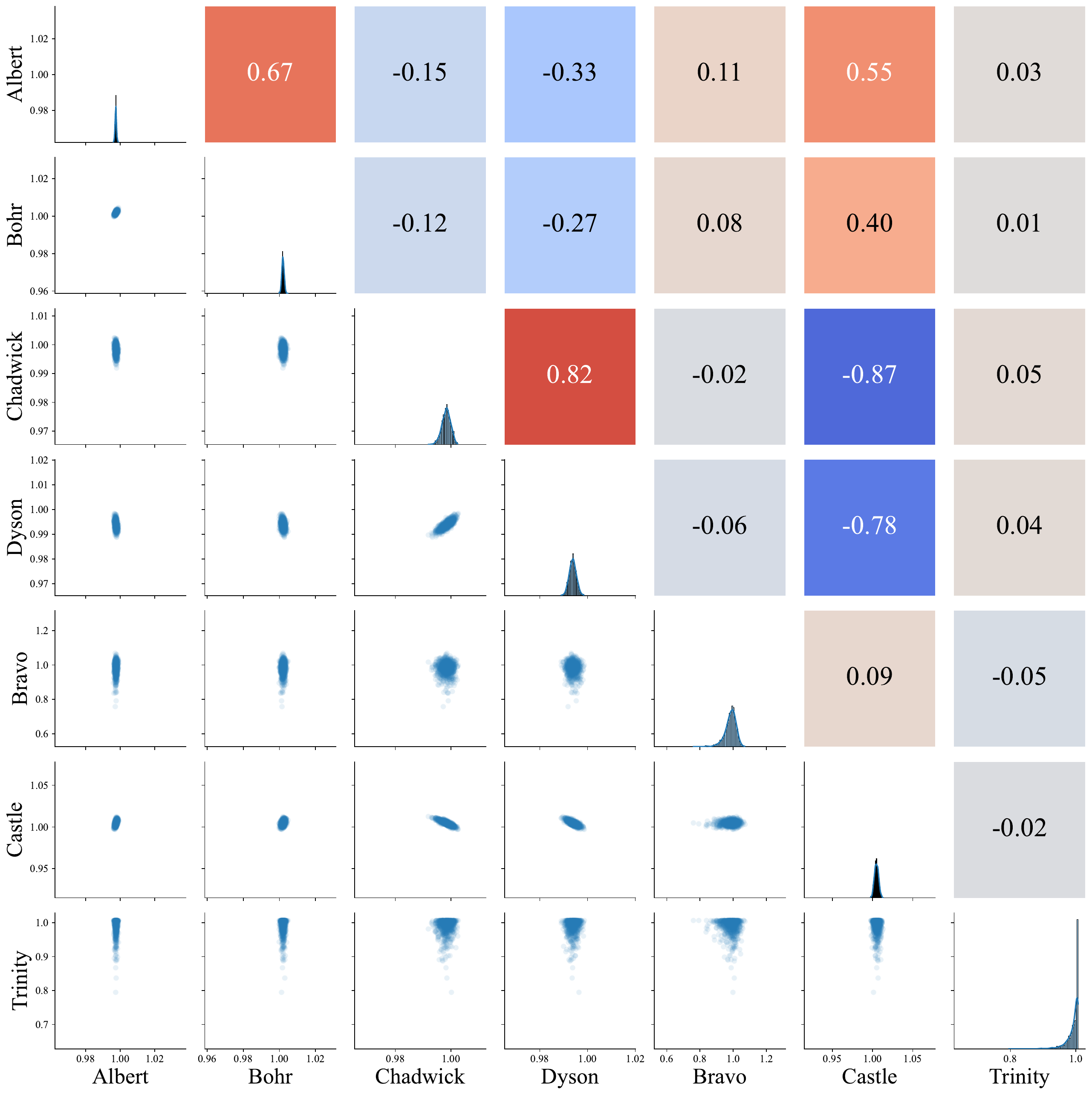}
        \caption{\centering Case 5: Predicted}
    \end{subfigure}
    \begin{subfigure}{0.32\linewidth}
        \centering
        \includegraphics[width=\linewidth]{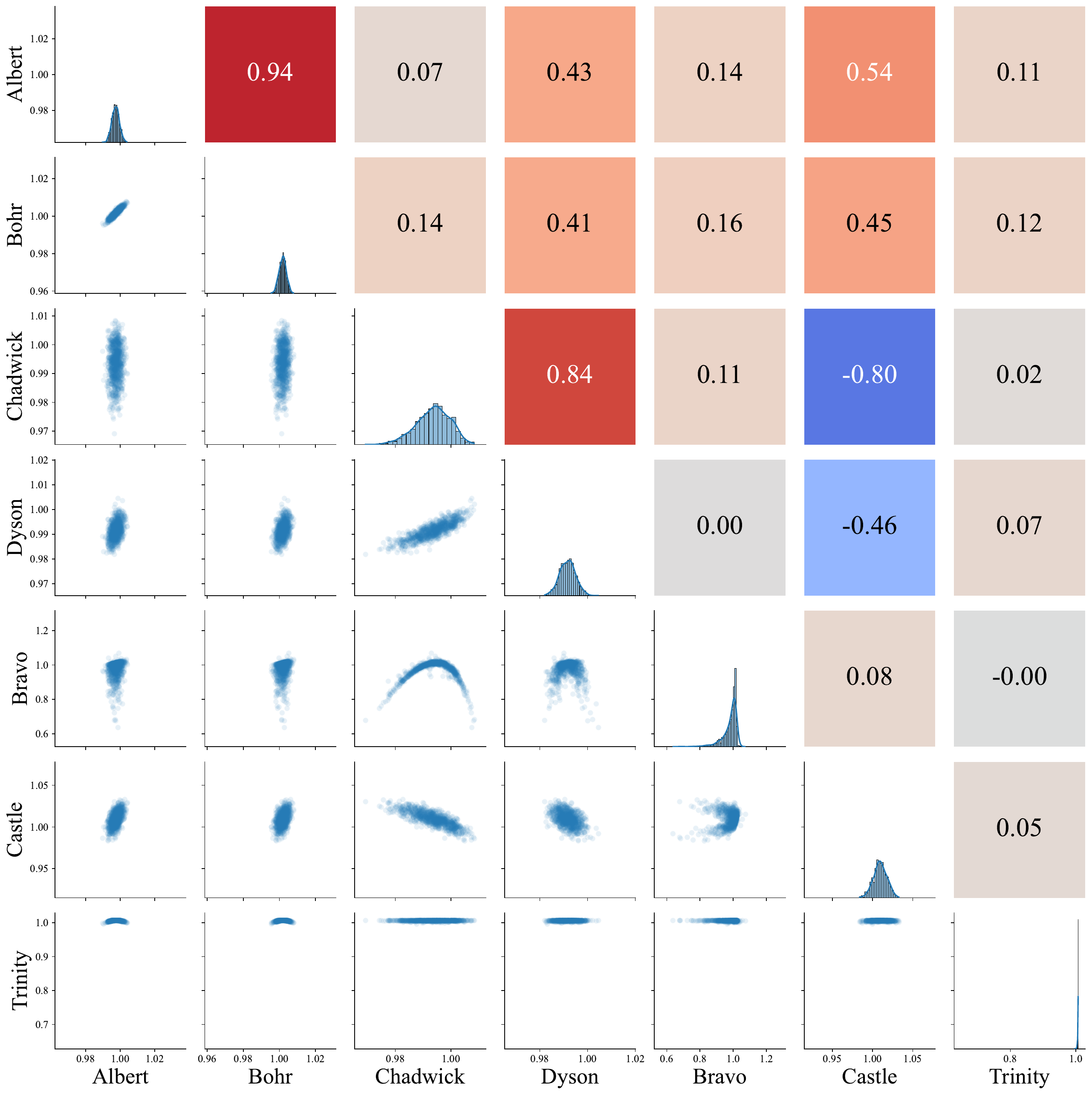}
        \caption{\centering Case 1: Computed}
    \end{subfigure}
    \begin{subfigure}{0.32\linewidth}
        \centering
        \includegraphics[width=\linewidth]{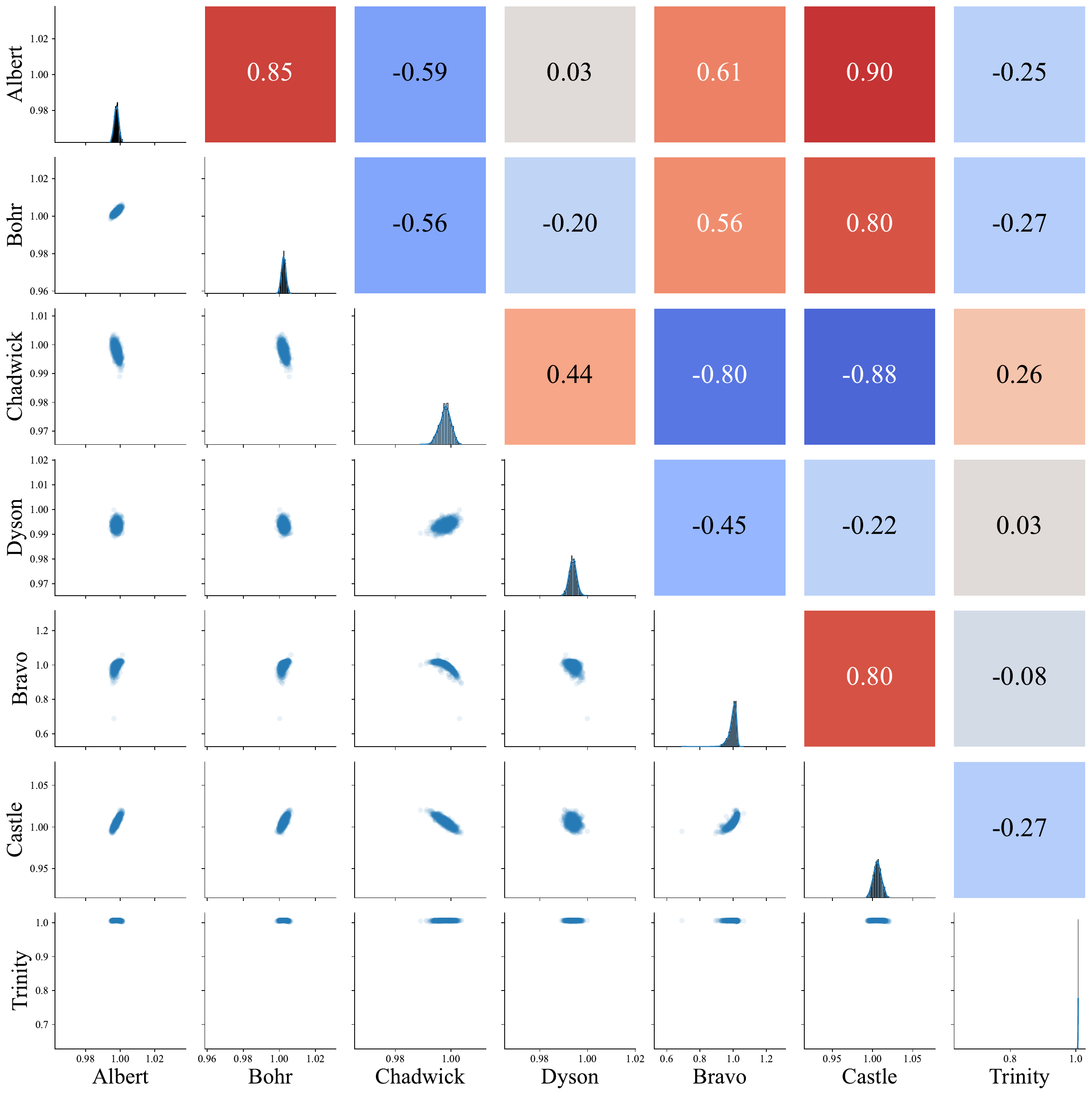}
        \caption{\centering Case 3: Computed}
    \end{subfigure}
    \begin{subfigure}{0.329\linewidth}
        \centering
        \includegraphics[width=\linewidth]{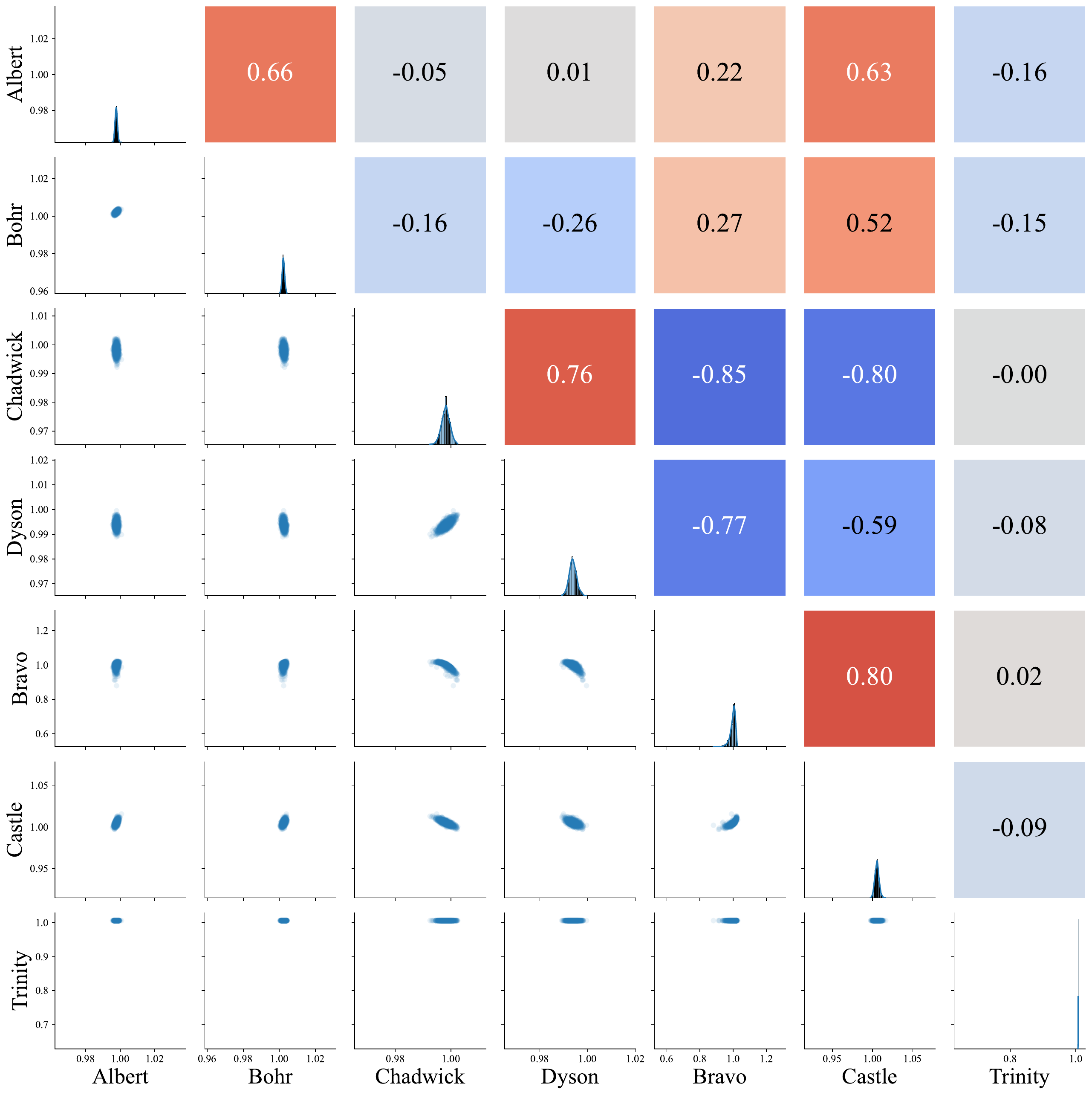}
        \caption{\centering Case 5: Computed}
    \end{subfigure}
    \caption{Comparison of predicted and computed MOCABA posterior distributions.}
    \label{fig:compare_mocaba}
\end{figure}

Pairwise plots for samples from the posterior predictive distributions and computed predictions at posterior parameter samples are shown for MOCABA in Fig.~\ref{fig:compare_mocaba}. Similar to GLLS, MOCABA has agreement between posterior predictions and computed responses for linear \textsc{Applications}. In contrast with GLLS, MOCABA is able to account for higher moments and produces skewed posterior predictive distributions similar to those computed for the nonlinear \textsc{Applications} \texttt{Bravo} and \texttt{Trinity}. While MOCABA does capture the skewed distributions, some distortion from the data transformation remains and can be seen when comparing \texttt{Bravo}'s and \texttt{Trinity}'s histograms and scatterplots. While posterior predictions fail to capture the full detail of relationships available in the computed scatterplots, this is not strictly necessary to quantify the posterior \textsc{Application} uncertainties as this in primarily reliant on the distribution scale.

\begin{figure}[!htb]
    \centering
    \begin{subfigure}{0.32\linewidth}
        \centering
        \includegraphics[width=\linewidth]{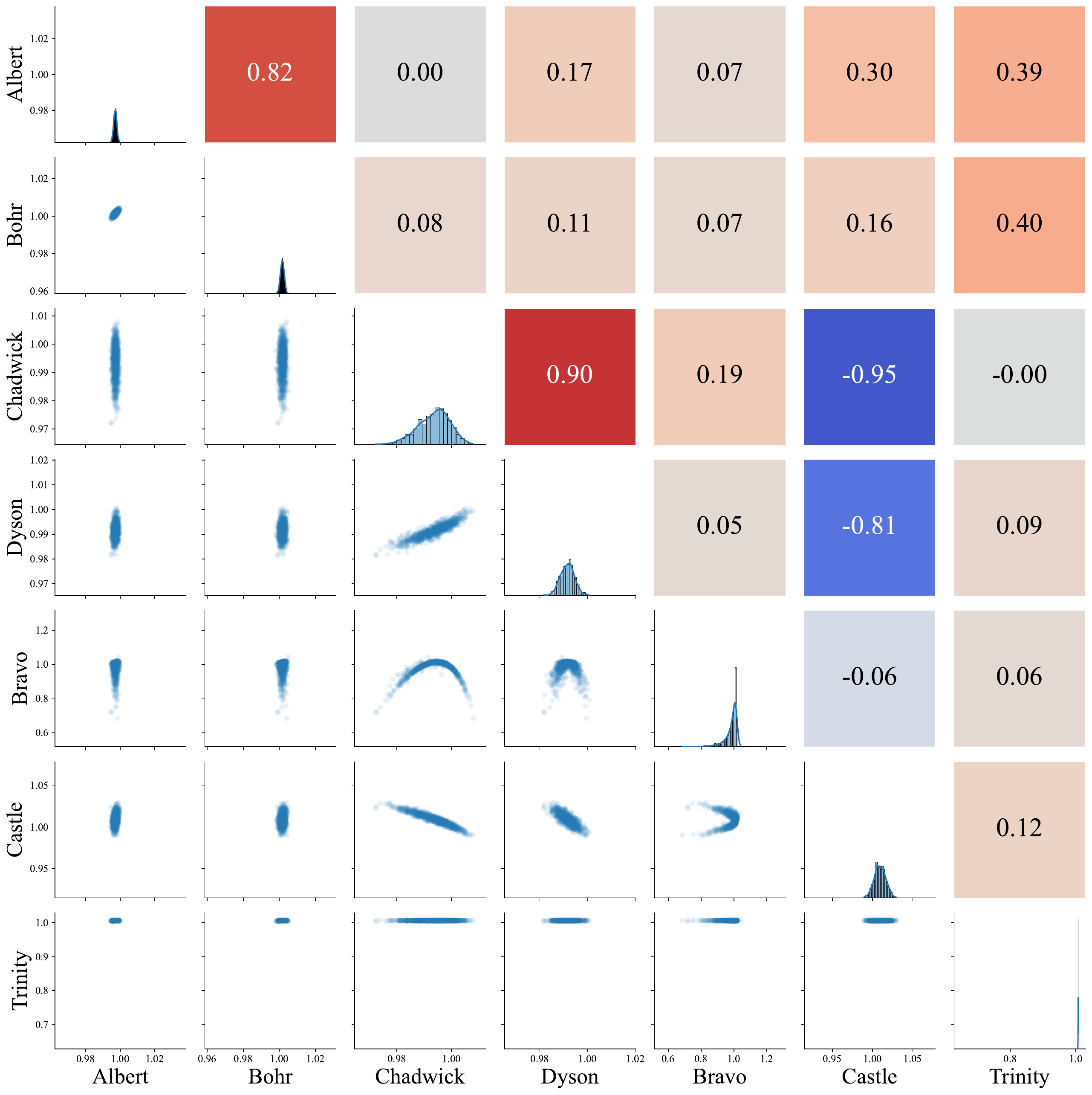}
        \caption{\centering Case 1: \texttt{Albert} \\ IUQ without $\delta$}
    \end{subfigure}
    \begin{subfigure}{0.32\linewidth}
        \centering
        \includegraphics[width=\linewidth]{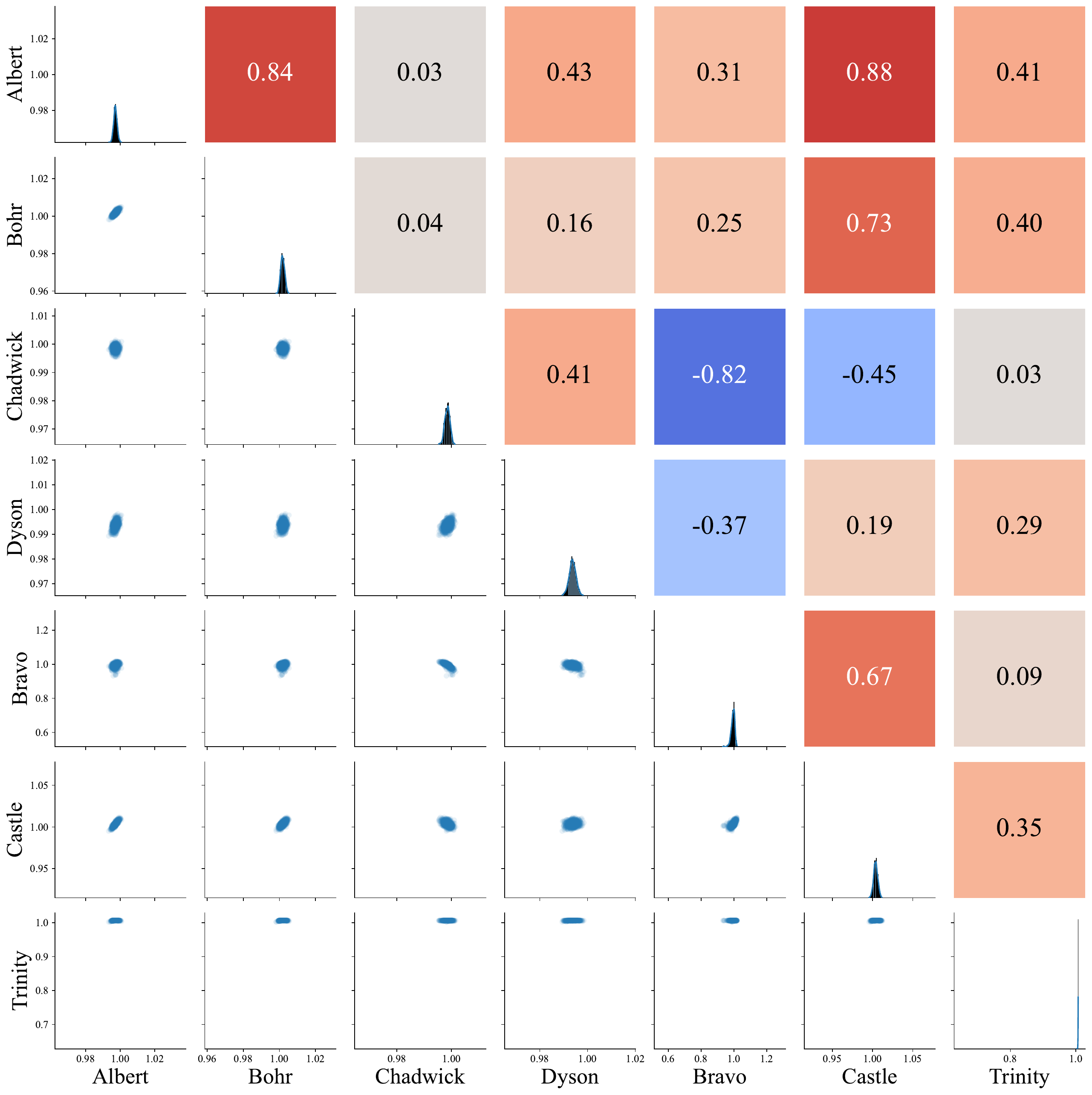}
        \caption{\centering Case 3: \texttt{Albert} + \texttt{Chadwick} \\ IUQ without $\delta$}
    \end{subfigure}
    \begin{subfigure}{0.32\linewidth}
        \centering
        \includegraphics[width=\linewidth]{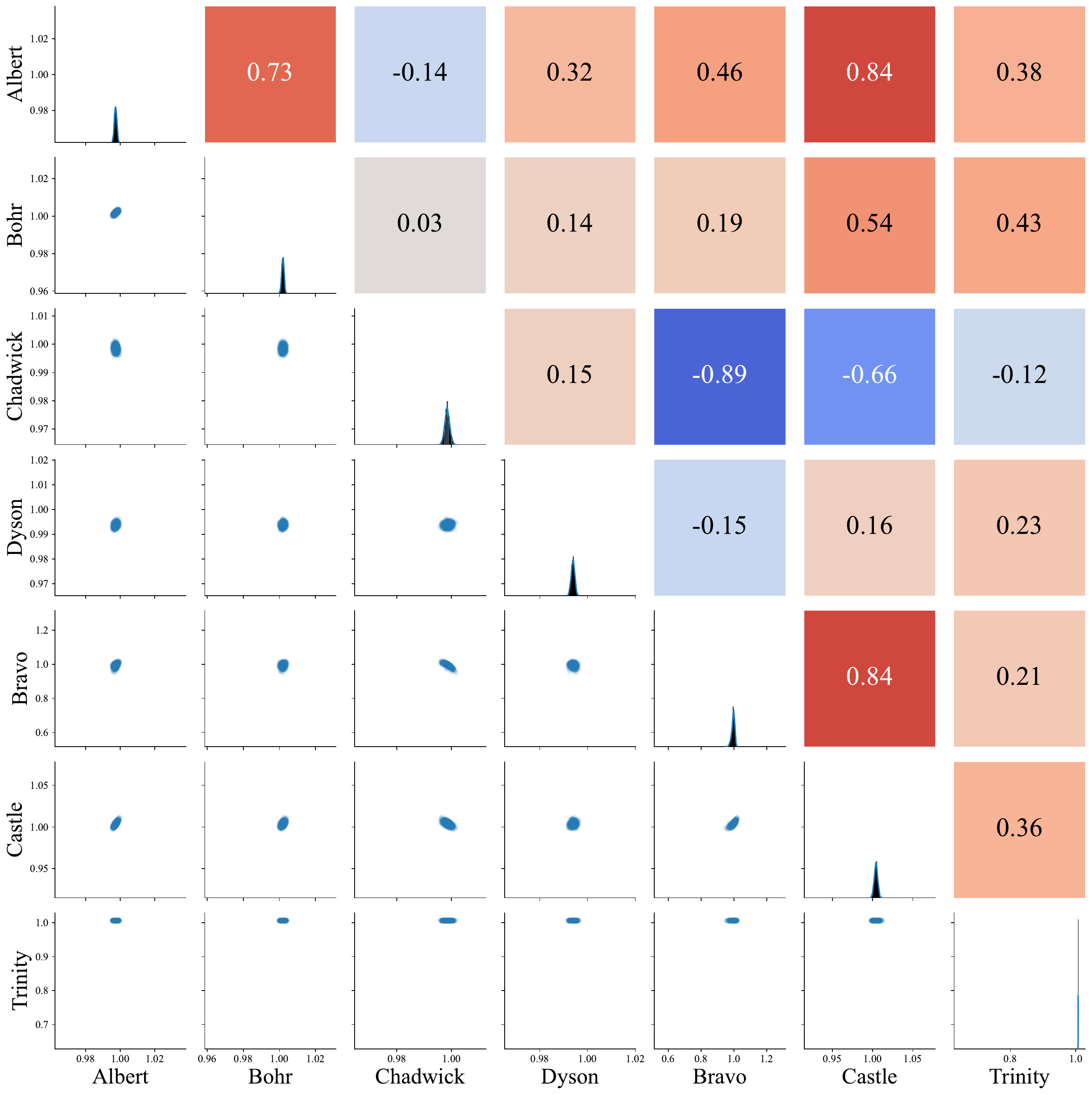}
        \caption{\centering  Case 5: All \\ IUQ without $\delta$}
    \end{subfigure}
    \begin{subfigure}{0.32\linewidth}
        \centering
        \includegraphics[width=\linewidth]{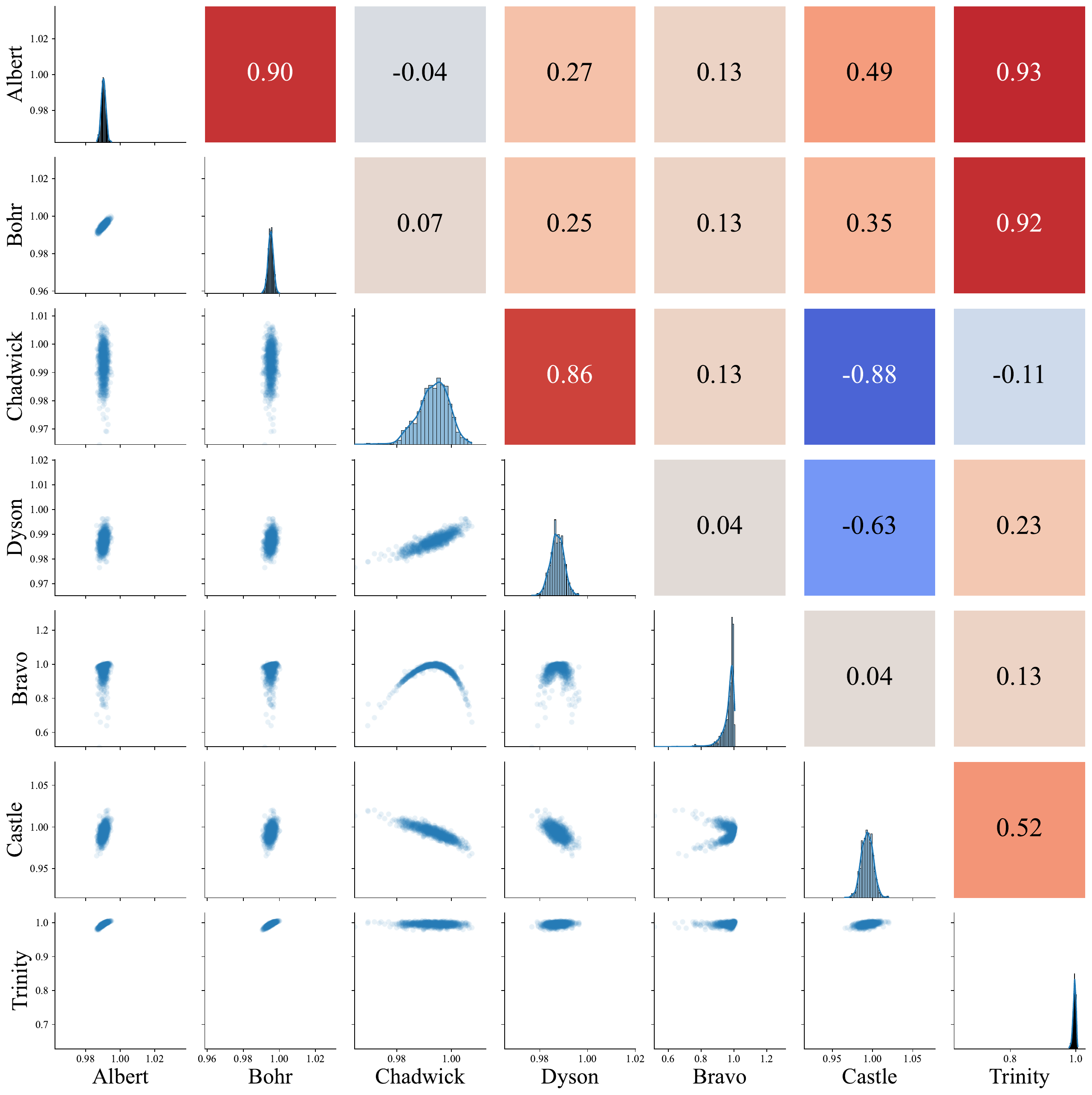}
        \caption{\centering Case 1: \texttt{Albert} \\ IUQ with $\delta$}
    \end{subfigure}
    \begin{subfigure}{0.32\linewidth}
        \centering
        \includegraphics[width=\linewidth]{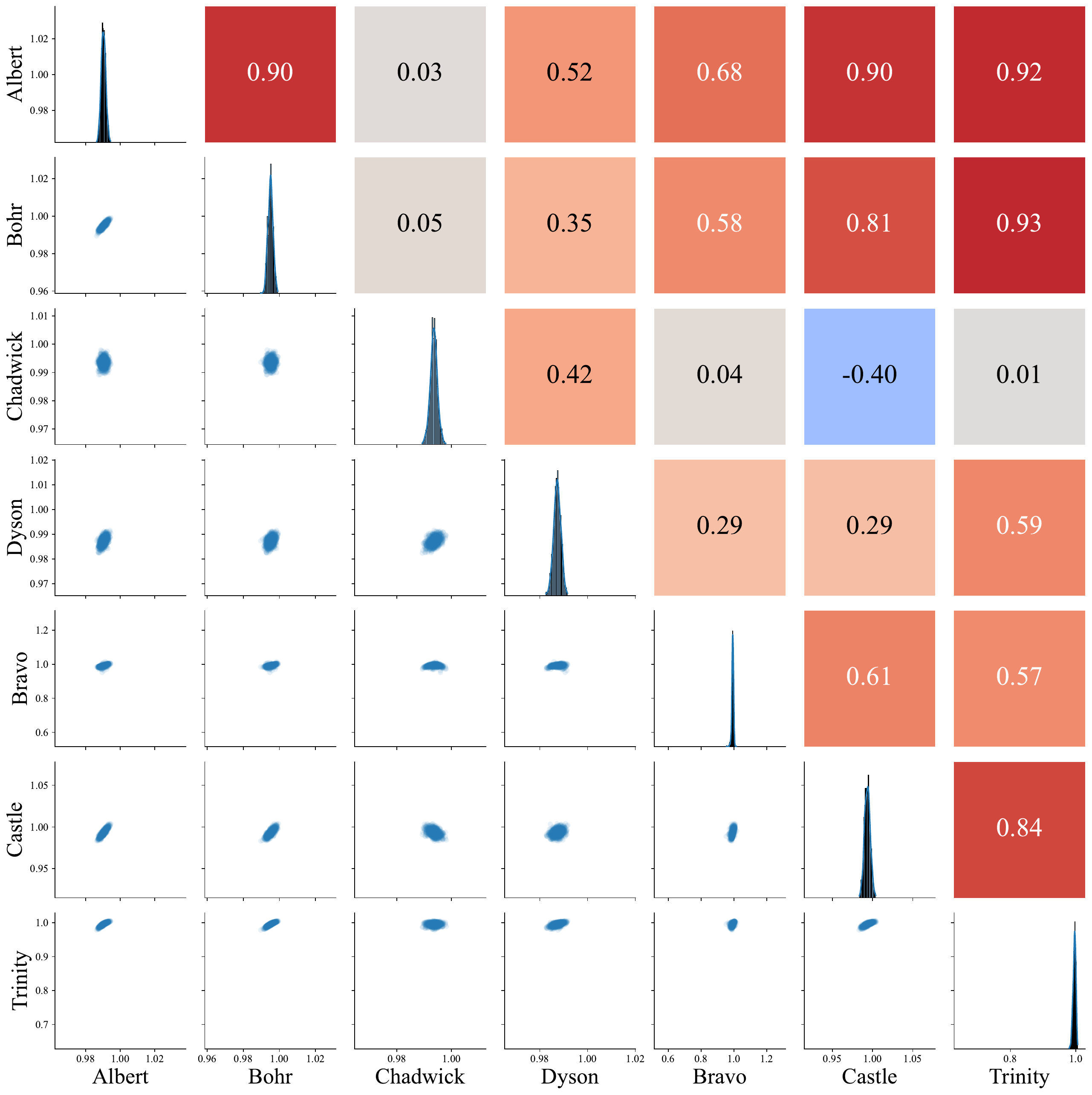}
        \caption{\centering Case 3: \texttt{Albert} + \texttt{Chadwick} \\ IUQ with $\delta$}
    \end{subfigure}
    \begin{subfigure}{0.329\linewidth}
        \centering
        \includegraphics[width=\linewidth]{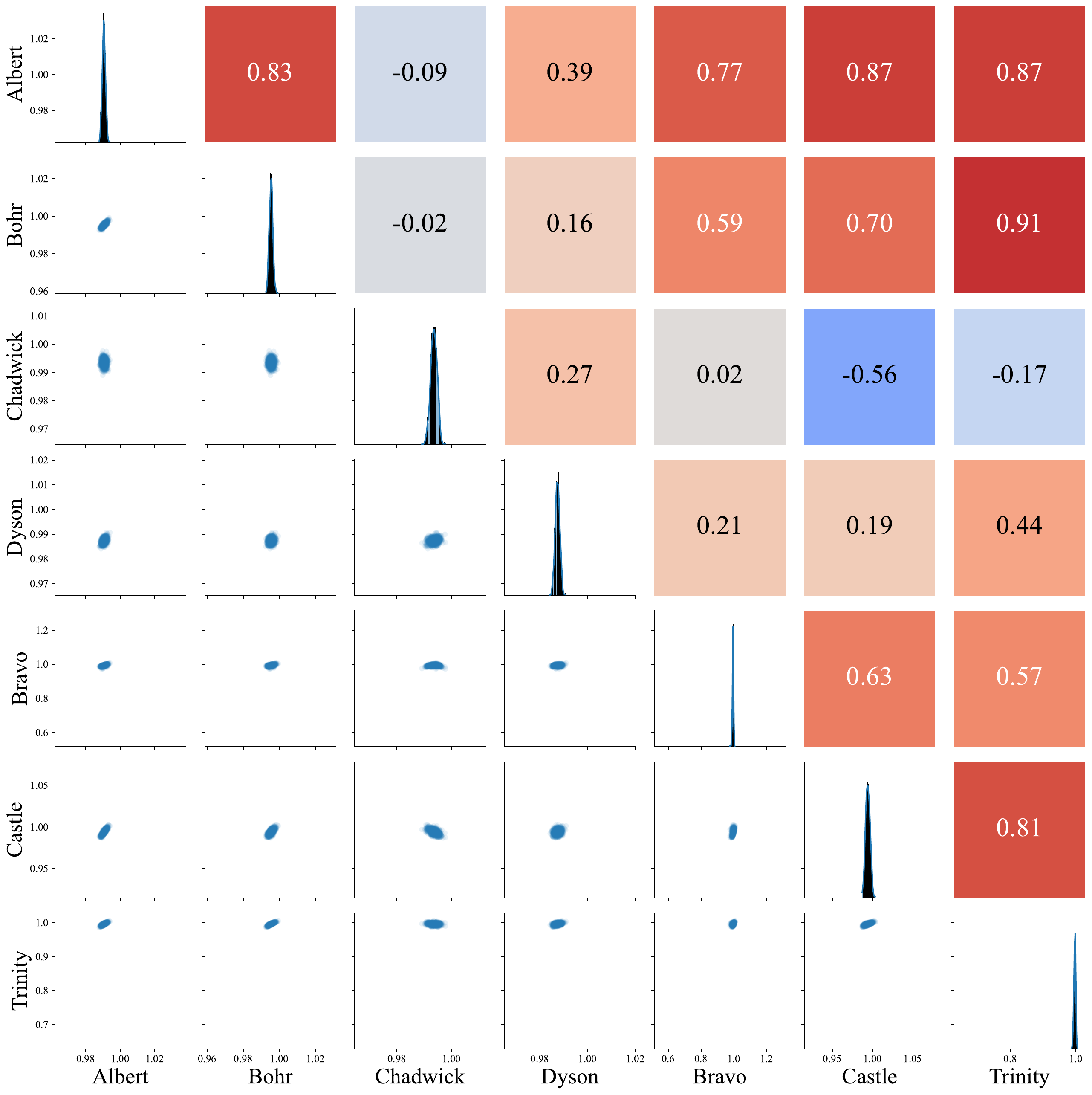}
        \caption{\centering Case 5: All \\ IUQ with $\delta$}
    \end{subfigure}
    \caption{Comparison of posterior distributions for IUQ with and without model discrepancy $\delta$.}
    \label{fig:compare_iuq}
\end{figure}

Since IUQ uses model evaluations at posterior parameter samples to produce the posterior predictive distributions, there is no difference between posterior predictions and computed distributions. Instead, pairwise posterior predictive distributions are shown for IUQ with and without and with model discrepancy $\delta$ considered in Fig.~\ref{fig:compare_iuq}. Inclusion of the model discrepancy term shifts the center and changes the scale of the posterior distributions. Some changes in the scale or variance of the posterior distributions can be attributed to additional variance introduced by the model discrepancy term itself. However, the model discrepancy term might shift the model to a region of the parameter domain with increased or decreased variance. This is demonstrated with \texttt{Bravo} in Cases 3 and 5, where the posterior predictive variance is significantly reduced when the model discrepancy term is included. Thus, the effect of model discrepancy on the posterior prediction variance is difficult to predict before analysis.

\subsection{IUQ Observations}
\label{subsection:IUQ-Observations}

Comparing posterior distributions of the cases provides valuable insight about the information extracted during analysis. Due to similarities in several of the \textsc{Experiment} model sensitivities, three case comparisons are particularly interesting and will be presented here. Pairwise plots for these comparisons occur on the same scale to enable easy comparisons.

\begin{figure}[!htb]
    \centering
    \begin{subfigure}{0.32\linewidth}
        \centering
        \includegraphics[width=\linewidth]{fig_SG14_pairwise_all_prior.pdf}
        \caption{\centering Prior}
    \end{subfigure}
    \begin{subfigure}{0.32\linewidth}
        \centering
        \includegraphics[width=\linewidth]{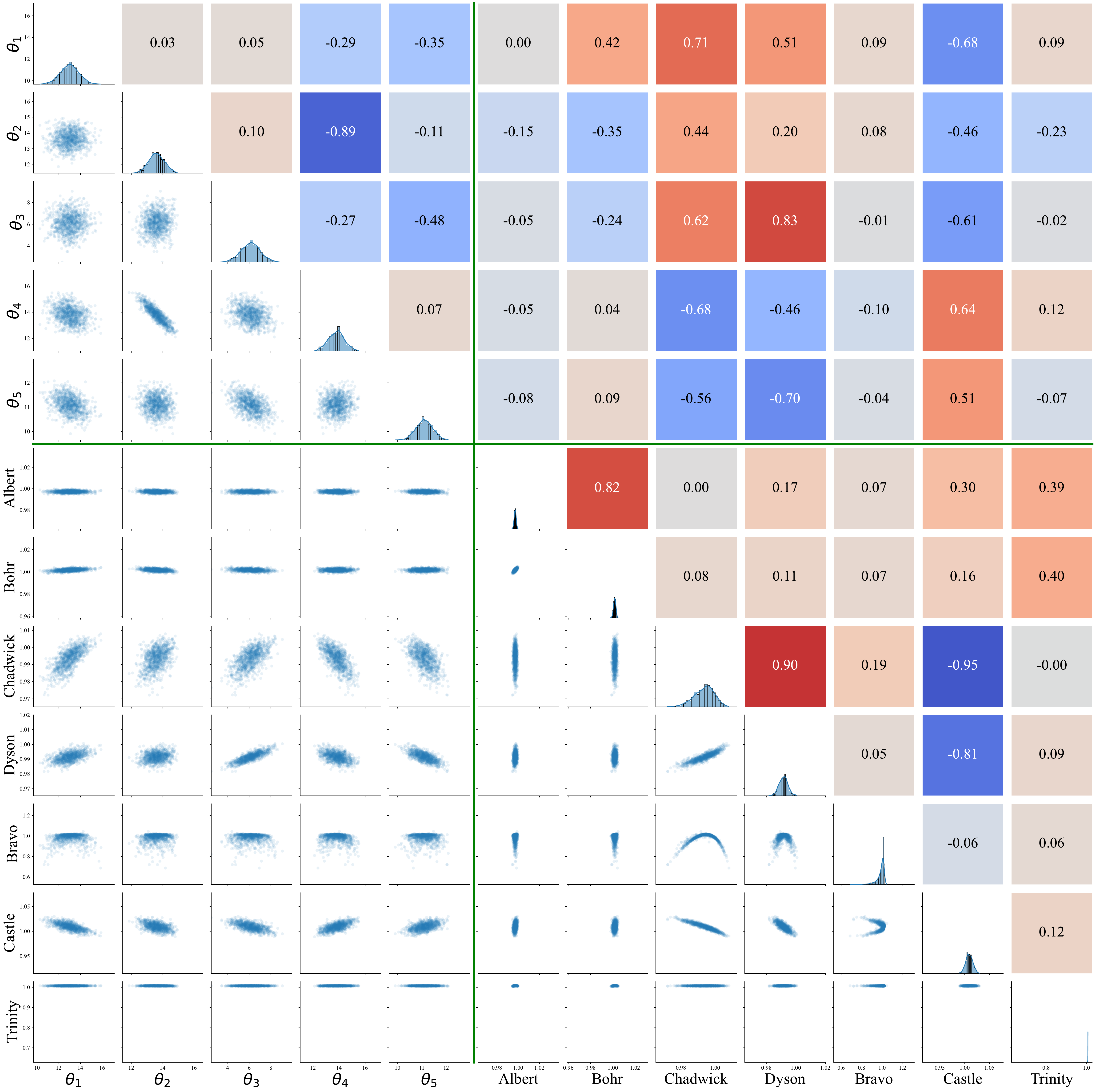}
        \caption{\centering Case 1: \texttt{Albert}}
    \end{subfigure}
    \begin{subfigure}{0.32\linewidth}
        \centering
        \includegraphics[width=\linewidth]{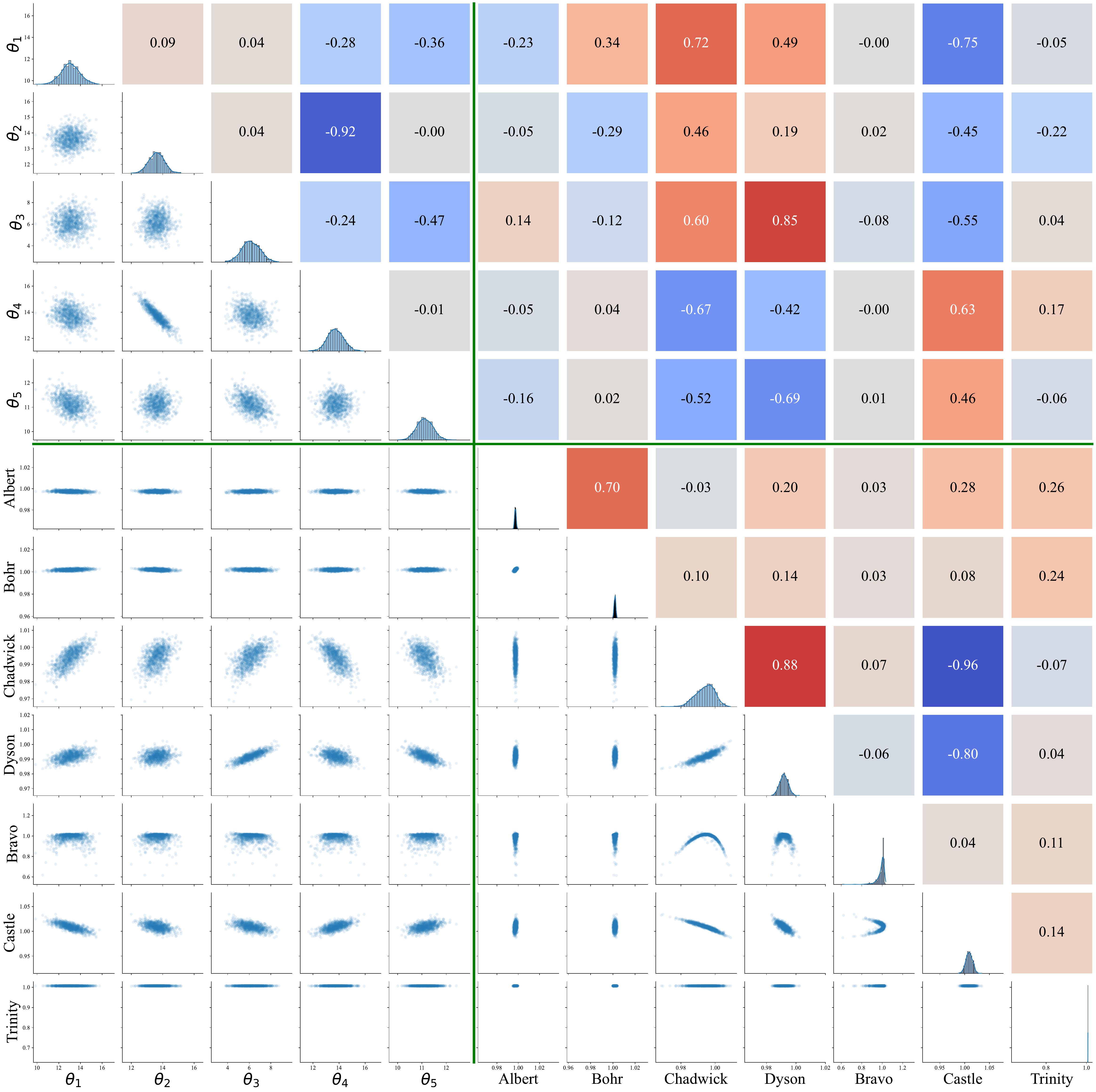}
        \caption{\centering Case 2: \texttt{Albert} + \texttt{Bohr}}
    \end{subfigure}
    \caption{Comparison of prior and posterior distributions for Case 1 and Case 2.}
    \label{fig:compare_prior12}
\end{figure}

Comparing the prior distribution with posterior distributions from Case 1 and Case 2 in Fig.~\ref{fig:compare_prior12} yields two important observations. The first observation is that \textsc{Applications} sharing strong dependence with the \textsc{Experiment} in the prior showed the most significantly reduced posterior distributions. This is particularly evident in \texttt{Bohr} and \texttt{Trinity}. Initially, this agrees with the conventional practice of favoring \textsc{Experiments} with high $c_k$ values indicating strong linear dependence. This is also immediately intuitive when examining the scatterplots and considering the effects of the MCMC process. MCMC will accept those parameters producing model evaluations within the measured data distribution and reject those outside the data distribution with high probability. In Case 1, which has only one data point, this means the posterior predictive distribution of \texttt{Albert} should resemble that of the data and uncertainty, which it does. In the Case 1 column of scatterplots of \texttt{Albert} with respect to each of the other models, this has the effect of ``trimming'' the tails of the corresponding scatterplots from the prior distribution or taking a ``slice'' through their center. Thus, those \textsc{Applications} with strong dependence with \texttt{Albert}, now have very narrow distribution in the Case 1 posterior. This also has the result of ``flattening'' \texttt{Albert}'s sensitivity with respect to each of the parameters as seen in the respective scatterplots. Thus, \texttt{Albert} now has relatively low sensitivity to the parameters over the domain of the posterior parameter distributions. Similarly, the sensitivities of \textsc{Application} responses with strong dependence with \texttt{Albert} are ``flattened'' as well.

The second observation is the posterior distributions have very little change from Case 1 to Case 2. This implies that the inclusion of data from \texttt{Bohr} does not provide significant information beyond what is already provided by \texttt{Albert}. The posterior predictive distribution of \texttt{Bohr} in Case 1 is already very near the data distribution. Using the previous metaphor, there is very little data to ``trim'' or ``slice'' in the scatterplots of \texttt{Bohr} with respect to the other models. This is accompanied by very ``flat'' sensitivities to the parameters over the adjusted parameter domain. The result is very little information is gained and the addition of \texttt{Bohr} will not serve to further reduce the variance in the posterior predictive distributions. It is important to note, inclusion of \texttt{Bohr} produces only a minor adjustment in its own posterior predictive distribution from Case 1 to Case 2 despite having perfect dependence with itself. Instead of dependence, it is the flat sensitivity profile that is a better indication \texttt{Bohr}'s inclusion will not provide significant information to the data adjustment. An exception to this occurs if the measurement uncertainty associated with \texttt{Bohr}'s data was significantly lower than \texttt{Albert}, then its inclusion would serve to further reduce variance in the posterior as it could further ``trim'' data in these scatter plots. Conversely, if \texttt{Bohr}'s posterior predictions in Case 1 did not align with the data provided by \texttt{Bohr}, inclusion of \texttt{Bohr} could serve to expand the parameter distribution and thus expand the variance in the posterior distributions for Case 2. This would be the case if data from \texttt{Bohr} provided evidence of parameter values that disagrees with the data from \texttt{Albert}. This implies that inclusion of \textsc{Experiments} with very strong dependence and very similar sensitivity profiles will not necessarily serve to further reduce uncertainties in the \textsc{Application}, unless the measured data has relatively lower uncertainty. However, their inclusion will make the resulting data adjustments more robust to errors in the measurement data.

\begin{figure}[!htb]
    \centering
    \begin{subfigure}{0.32\linewidth}
        \centering
        \includegraphics[width=\linewidth]{fig_SG14_pairwise_all_prior.pdf}
        \caption{\centering Prior}
    \end{subfigure}
    \begin{subfigure}{0.32\linewidth}
        \centering
        \includegraphics[width=\linewidth]{fig_SG14_pairwise_all_iuq1.pdf}
        \caption{\centering Case 1: \texttt{Albert}}
    \end{subfigure}
    \begin{subfigure}{0.32\linewidth}
        \centering
        \includegraphics[width=\linewidth]{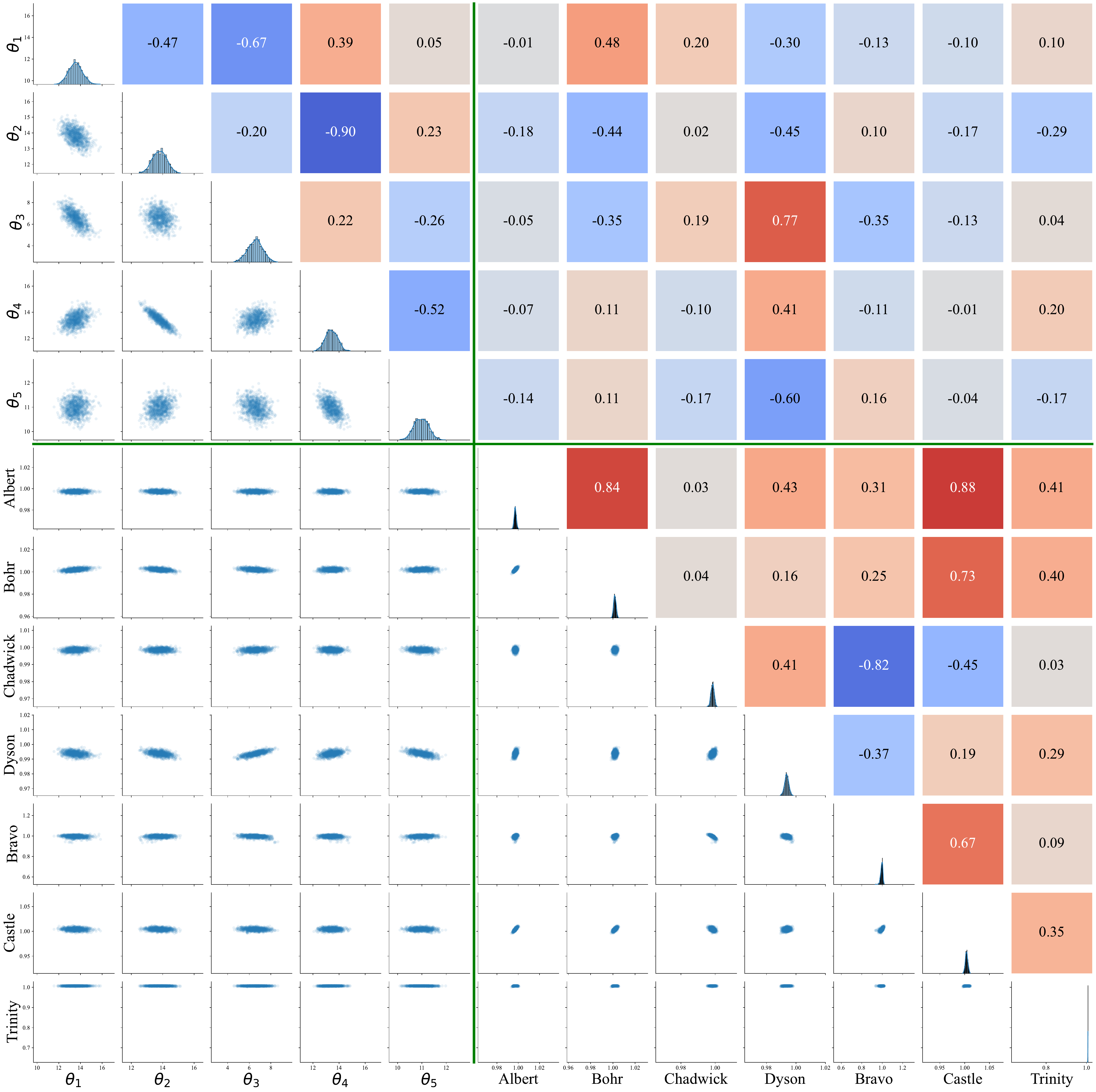}
        \caption{\centering Case 3: \texttt{Albert} + \texttt{Chadwick}}
    \end{subfigure}
    \caption{Comparison of prior and posterior distributions for Case 1 and Case 3.}
    \label{fig:compare_prior13}
\end{figure}

The next comparison involves the prior, Case 1, and Case 3 shown in Fig.~\ref{fig:compare_prior13}. Here we observe significant information gain from Case 1 to Case 3 indicated by reduced variance in the posterior parameter and predictive distributions. It is noted from the prior scatterplots that \texttt{Chadwick} has low dependence with most other models except \texttt{Bravo} and maybe \texttt{Trinity}. Additionally, using the conventional metric, it has a low correlation $c_k$ value with the other models, and even anti-correlation with \texttt{Castle}, which is also a linear model. This result appears counter to the conventional practice of excluding low correlation \textsc{Experiments} from data adjustments. However, it is clear from the scatterplots why \texttt{Chadwick} remains informative to the data adjustment. While \texttt{Chadwick} generally has low dependence with other models in the prior, it is revealed in the Case 1 posterior that in this adjusted subset of the parameter domain, \texttt{Chadwick} has much stronger dependence with other models, especially noted in the \texttt{Dyson} and \texttt{Castle} scatter plots with \texttt{Chadwick}. More importantly, in \texttt{Chadwick}'s scatterplots with respect to the parameters, the sensitivities have not been ``flattened,'' indicating the model will provide additional information to the data adjustment. Similar to the previous comparison, in transitioning from Case 1 to Case 3 we observe a ``trimming'' of the posterior predictive distribution in \texttt{Chadwick} to resemble the data distribution. This is accompanied by the subsequent variance reductions in \textsc{Applications} with strong dependence and ``flattening'' of \texttt{Chadwick}'s sensitivity profile.

\begin{figure}[!htb]
    \centering
    \begin{subfigure}{0.32\linewidth}
        \centering
        \includegraphics[width=\linewidth]{fig_SG14_pairwise_all_prior.pdf}
        \caption{\centering Prior}
    \end{subfigure}
    \begin{subfigure}{0.32\linewidth}
        \centering
        \includegraphics[width=\linewidth]{fig_SG14_pairwise_all_iuq3.pdf}
        \caption{\centering Case 1: \texttt{Albert}}
    \end{subfigure}
    \begin{subfigure}{0.32\linewidth}
        \centering
        \includegraphics[width=\linewidth]{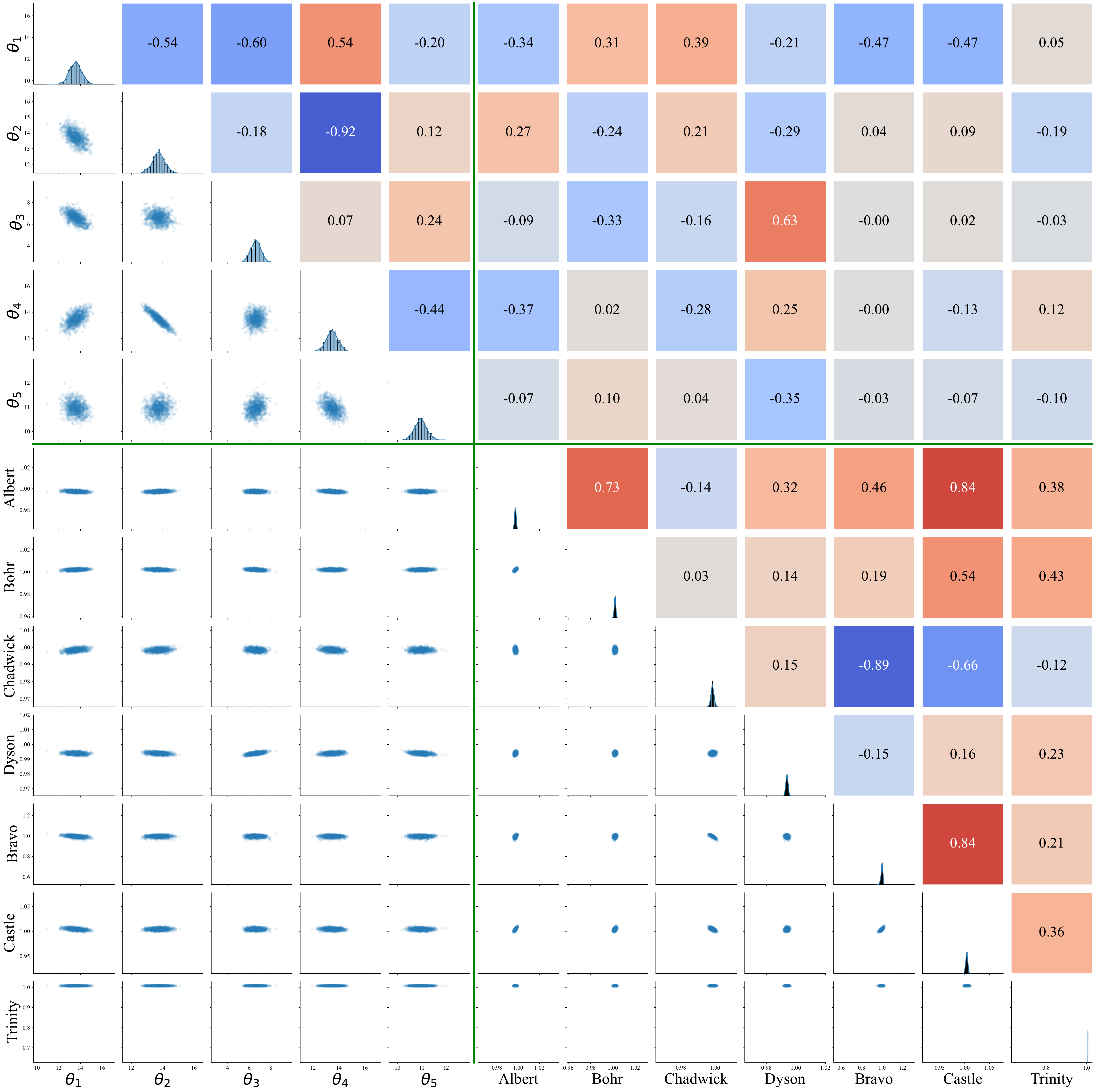}
        \caption{\centering Case 2: \texttt{Albert} + \texttt{Bohr}}
    \end{subfigure}
    \caption{Comparison of prior and posterior distributions for Case 3 and Case 5.}
    \label{fig:compare_prior35}
\end{figure}

The last comparison involves the prior, Case 3, and Case 5 shown in Fig.~\ref{fig:compare_prior35}. As with the first comparison, only minor adjustments appear in the posterior distributions from Case 3 to Case 5. In the prior, \texttt{Albert}, \texttt{Bohr}, and \texttt{Dyson} share very similar sensitivity profiles as indicated by the scatterplots with respect to the parameters. They also share relatively similar model response behaviors, as indicated by their linear dependence in the scatterplots of the models with respect to each other. This comparison reinforces the analysis of the first comparison. 

These observations imply that a collection of informative \textsc{Experiments} will include those with minimum measurement uncertainty and differing sensitivity profiles for parameters shared with the \textsc{Application}. While strong dependence between an \textsc{Experiment} and \textsc{Application} might indicate the experiment is informative to the data adjustment, it is not a guarantee. Similarly, weak dependence does not necessarily indicate an \textsc{Experiment} should be excluded from a data adjustment. As seen in the comparison of Case 1 and Case 3, there may be strong dependence between \textsc{Experiment} and \textsc{Application} within a subset of the parameter domain that is not immediately apparent. While including multiple \textsc{Experiments} with similar sensitivity profiles may not serve to significantly reduce the posterior predictive variances, their inclusion will contribute to the overall robustness of the data adjustment.

\section{Summary \& Conclusions}
\label{section:Conclusions}

Data adjustments are performed on a set of four \textsc{Experiments} and three \textsc{Applications} as part of a performance benchmark exercise to assess data adjustment methods for nonlinear applications. The performance benchmark directs analysis of five comparison cases, each using a subset of the experimental data. GLLS, MOCABA, and IUQ are performed and their results compared.

Observations from this performance benchmark exercise indicate that experiment sensitivity profile comparisons reveal more about the quantity of information a set of experiments provides during data adjustment than correlation comparisons. Additionally, a correlation comparison quickly looses its value as applications exhibit increased nonlinear behaviors. Other measures of nonlinear dependence may mislead analysts to believe an experiment will be uninformative due to lack of dependence with the application, when there is dependence hidden within a subset of the parameter domain. Typical methods of sensitivity computation remain valid when using experiments with linear behaviors. These sensitivities could be used as an assessment metric for inclusion in a set of measured data to be used in a data adjustment. While inclusion of experiments with similar sensitivities does not tend to further reduce posterior prediction variances, it contributes to a more robust data adjustment.

Performance of GLLS is unsurprisingly reaffirmed for linear applications in this exercise, however its underlying assumptions make it ill-suited for nonlinear applications. MOCABA and IUQ perform comparably in this benchmark exercise. MOCABA addresses nonlinearities through a lightweight sampling and distribution transformation with only slight loss of resolution in the posteriors. IUQ uses MCMC to sample directly from the posterior, but may suffer when scaling to larger parameter dimension.

\section*{Acknowledgments}

This material is based upon work supported by the Consortium for Nuclear Forensics (CNF) under the U.S. Department of Energy (DOE) National Nuclear Security Administration (NNSA) award number DE-NA0004142. Any opinions, findings, and conclusions or recommendations expressed in this paper are those of the authors and do not necessarily reflect the views of the U.S. DOE.

\pagebreak
\bibliographystyle{ans_js}
\bibliography{bibliography.bib}

\end{document}